\documentclass[useAMS,usenatbib]{mn2e}

\usepackage{graphicx}
\usepackage{txfonts}
\usepackage[usenames,dvipsnames]{color}

\usepackage{epsfig}

\def\e10{\eta_{10}}

\def\etal{et al.}
\def\iso#1#2{\mbox{${}^{#2}{\rm #1}$}}

\def\b1#1{\iso{B}{1#1}}

\def\beq{\begin{equation}}
\def\eeq{\end{equation}}
\def\beqar{\begin{eqnarray}}
\def\eeqar{\end{eqnarray}}
\def\simlt{\lower.5ex\hbox{$\; \buildrel < \over \sim \;$}}
\def\simgt{\lower.5ex\hbox{$\; \buildrel > \over \sim \;$}}
\def\simpropto{\lower.2ex\hbox{$\; \buildrel \propto \over \sim \;$}}

\newcommand{\apj}{ApJ}
\newcommand{\apjs}{ApJS}
\newcommand{\apjl}{ApJL}

\newcommand{\aap}{A\&A}
\newcommand{\nat}{Nat}
\newcommand{\aj}{AJ}
\newcommand{\nphysa}{NPhysA}
\newcommand{\mnras}{MNRAS}
\newcommand{\procspie}{Proc. SPIE}
\newcommand{\physrep}{PhysRep}
\newcommand{\pasp}{PASP}

\newcommand{\araa}{Ann. Rev. Astron. Astroph.}

\title[Cosmic NSM Rate and GW constrained by  the R Process]{Cosmic Neutron Star Merger Rate and Gravitational Waves constrained by  the R Process Nucleosynthesis}
 
\author[E. Vangioni, S. Goriely, F.  Daigne, P. Fran\c cois, K. Belczynski]
{Elisabeth Vangioni$^{1}$\thanks{e-mail:vangioni@iap.fr}, 
St\'ephane Goriely$^{2}$, 
Fr\'ed\'eric Daigne$^{1}$, 
Patrick Fran\c cois$^{3}$, \newauthor and Krzysztof Belczynski$^{4}$\\
$^{1}$Sorbonne Universit\'es, UPMC Univ Paris 6 et CNRS, UMR 7095, Institut d'Astrophysique de Paris, 98 bis bd Arago, 75014 Paris, France\\
$^{2}$Institut d'Astronomie et d'Astrophysique, CP 226,  Universit\'e Libre de Bruxelles, 1050 Brussels, Belgium\\
$^{3}$GEPI, Paris-Meudon Observatory, 61 Avenue de l'Observatoire, F-75014 Paris, France\\
$^{4}$Astronomical Observatory, University of Warsaw, Al. Ujazdowskie 4, 00-478 Warsaw, Poland
}
 
\begin{document}

\pagerange{\pageref{firstpage}--\pageref{lastpage}} \pubyear{2015}
\maketitle
\label{firstpage}

\begin{abstract}
The cosmic evolution of the neutron star merger (NSM) rate can be deduced from the observed cosmic star formation rate. This allows to estimate the rate expected in the horizon of the gravitational wave detectors advanced Virgo and ad LIGO and to compare those rates with independent predictions. In this context, the rapid neutron-capture process, or r-process, can be used as a constraint assuming NSM is the main astrophysical site for this nucleosynthetic process.  We compute the early cosmic evolution of a typical r-process element, Europium. Eu yields from NSM are taken from recent nucleosynthesis calculations. The same approach allows to compute the cosmic rate of  Core Collapse SuperNovae (CCSN) and the associated evolution of Eu.

We find that the bulk of  Eu observations at ${\rm [Fe/H]} > -2.5$  can be  rather well fitted by either CCSN or NSM scenarios. However, at lower metallicity, the early Eu cosmic evolution favors NSM as the main astrophysical site for the r-process. A comparison between our calculations and spectroscopic observations at very low metallicities allows us to constrain the coalescence timescale in  the NSM scenario to $\sim$ 0.1--0.2~Gyr. 
These values are in agreement with the coalescence timescales of some observed binary pulsars.
 Finally, the cosmic evolution of Eu is used to put constraints on i) the NSM rate, ii) the merger rate in the horizon of the gravitational wave detectors advanced Virgo/ad LIGO, as well as iii) the expected rate of 
 electromagnetic counterparts to mergers ("kilonovae") in large near-infrared surveys.

\end{abstract}


\section{Introduction}

Recently, a special attention has been paid to neutron star (NS)--NS or NS--black hole (BH) mergers (hereafter NSM),  considered today as the most promising sources of gravitational waves for ground-based detectors such as advanced Virgo/ad Ligo
\citep{phinney91,narayan91,virgo09, abadie10, ligovirgo11,Aasi14a, Aasi14b, Aasi14c, Bel14}
   but also because several evidences \citep[see][for a recent review]{berger14} suggest in addition that these events are 
the progenitors of short gamma-ray bursts (GRB)
\citep{paczynski86,eichler89,narayan92,mochkovitch93}. 
Recent simulations show that, after the merger, the relic BH-torus system can produce an ultra-relativistic ejection along its rotation axis \citep{rezzolla11}, 
potentially leading to a short GRB. The similarities between the properties of the prompt emission in short and long GRBs \citep[se e.g.][]{guiriec10} suggest that the same dissipation process is at work in these two classes of bursts \citep{bosnjak14} and that the main differences are the lifetime of the central engine, leading in the case of binary compact objects to shorter events, and the density of the circumburst medium, leading to weaker afterglows \citep{nakar07}.
In addition,  hydrodynamic simulations have confirmed that
a non-negligible amount of matter, typically about $10^{-3}$ to $10^{-2}$~M$_\odot$, can be ejected 
quasi-isotropically
 \citep{janka99,ros99,ros04,oech07,gor11,baus13,gor13,just14,wa14} and that the matter ejected in such NSM events is essentially made of r-process nuclei, shedding a new light on a old mystery.
 
 The r-process, or the rapid neutron-capture process, of stellar nucleosynthesis
is invoked to explain the  production of the stable  (and some long-lived radioactive) neutron-rich nuclides 
heavier than iron that are observed in stars of various metallicities, as well as in the solar
system \citep[for a review, see][]{ar07}. The r-process remains the most complex nucleosynthetic process to model from 
the astrophysics as well as nuclear-physics points of view. The site(s) of
the r-process is (are) not identified yet,  all the proposed scenarios facing serious
problems. Complex---and often exotic---sites have been considered in the hope of
identifying astrophysical conditions in which the production of neutrons is large
enough to give rise to a successful r-process. 

Progress in the modelling of type-II
supernovae and long GRBs
 has raised a lot of excitement about the so-called
neutrino-driven wind environment. However, until now a successful r-process cannot be 
obtained {\it ab initio} without tuning the relevant parameters (neutron
excess, entropy, expansion timescale) in a way that is not supported by the most 
sophisticated existing models \citep{wanajo11,janka12}. 

Early in the development of the theory of nucleosynthesis, an alternative to the r-process 
in high-temperature supernova environments was proposed \citep{tsuruta65}.
It relies on the fact that at high densities (typically $\rho > 10^{10}$ g/cm$^{3}$) matter tends to be composed of nuclei lying on the neutron-rich side of the valley of nuclear stability as a result of free-electron
 captures. The astrophysical plausibility of this  scenario in accounting for the production of the r-nuclides has long been questioned.  It remained largely unexplored until the study  of the decompression of cold neutronised matter resulting from tidal effects of a BH on a NS companion \citep{lattimer74,lattimer77,meyer89}. 
Many investigations with growing sophistication have now confirmed NSM ejecta as viable sites  for strong r-processing~\citep{ar07,gor11,baus13,gor13,just14,frei99,go05,metzger10,roberts11,korobkin12,wa14}. In particular, recent nucleosynthesis calculations  \citep{just14} show that the combined contributions of both the dynamical (prompt) ejecta expelled during the binary NS-NS or NS-BH merger and the neutrino and viscously driven outflows generated during the post-merger remnant evolution of the relic BH-torus systems lead to the production of r-process elements from  $A \ga90$ up to thorium and uranium with an abundance distribution that reproduce extremely well the solar distribution, as well as the elemental distribution spectroscopically determined in very-low-metallicity stars.
The ejected mass of r-process material, combined with the predicted astrophysical event rate (around 
$10^{-5}$\,yr$^{-1}$ in the Milky Way \citep{dom12} can account for the majority of r-material in our Galaxy~\citep{gor11,baus13,just14}.

In NSM, nearly
 all of the ejecta are converted to r-process nuclei, whose radioactive
decay heating leads to potentially observable electromagnetic radiation in
the optical and infrared bands~\citep{li98,metzger10} with 100--1000 times fainter peak brightnesses than those of typical supernovae and durations of only days~\citep{gor11,roberts11,baus13}. These ``macronovae'' or ``kilonovae'' are intensely searched for (with a recent, possible first success in association with the short GRB~130603B, see~\citealt{berger13,tanvir13}). Their unambiguous discovery would constitute the first detection  of r-material in situ. 


In the present paper, our primary goal is to predict the cosmic evolution of the NSM that can be deduced from the observed history of the star formation rate (SFR), including the recent constraints obtained at high redshift \citep{Behroozi13, Behroozi14, O14, Bouwens14, Kistler13, Trenti13}. Computing the NSM rate requires to determine poorly known parameters, namely the fraction of NS in a binary system with another NS or a BH, and the characteristic coalescence timescale of such a system. 
 To constrain these parameters, we model the associated evolution of a typical r-process element, Europium, by assuming NSM as the main astrophysical site of the r-process and adopting the most recent nucleosynthetic calculations \citep{gor11,baus13,just14}, and we compare to the available observations in old stars of the Galactic halo and in external galaxies. Once the merger rate has been determined, it is possible to estimate the expected event rate for the new generation of gravitational wave detectors (advanced Virgo and LIGO), as well as the expected detection rate of kilonovae in present and future optical/near-infrared large surveys. The same approach can be followed to determine, without any new parameter, the CCSN rate, hence the predicted evolution of Europium in the alternative scenario where r-process elements are produced mainly in such stellar explosions.
 
The paper is organized as follows: in Sect.~\ref{sec:rates}, we select three possible scenarios for the SFR, which are representative of the current uncertainties at high redshift, and we describe how the CCSN and NSM rates can be deduced from the SFR, both scenarios still depending on given unknown parameters. In Sect.\ref{sect_cem}, we describe the chemical evolution model that we use to deduce from the cosmic SFR the expected evolution of the abundances of various chemical elements in star-forming structures. We show that the three scenarios considered in the paper for the cosmic SFR are compatible with a large set of observations, hence validating the present model. In \S~\ref{sect_site}, we show how the model can be extended to heavy elements produced by the r-process. In Sect.~\ref{sect_res}, we present the predicted cosmic evolution of Eu, assuming either the CCSN or the NSM as the main astrophysical site of the r-process. We make a detailed comparison with observations at low metallicities which correspond to the early evolution in the Universe and appear to be the most discriminative  between both sites. In the NSM case, this allows us in addition to constrain the properties of the binary system, in particular coalescence timescale. The resulting lower and upper limits on the merger rate are used in Sect.\ref{sec:mergerrates} to make predictions for the expected event rate in the horizon of advanced Virgo/LIGO, as well as for the kilonova rate in future surveys. Finally, we draw our conclusions in Sect.~\ref{sect_conc}.

\section{Predicting the NSM rate}
\label{sec:rates}
\subsection{The cosmic star formation rate (SFR)}
\label{sec:sfr}

\begin{figure}
\begin{center}
\epsfig{file=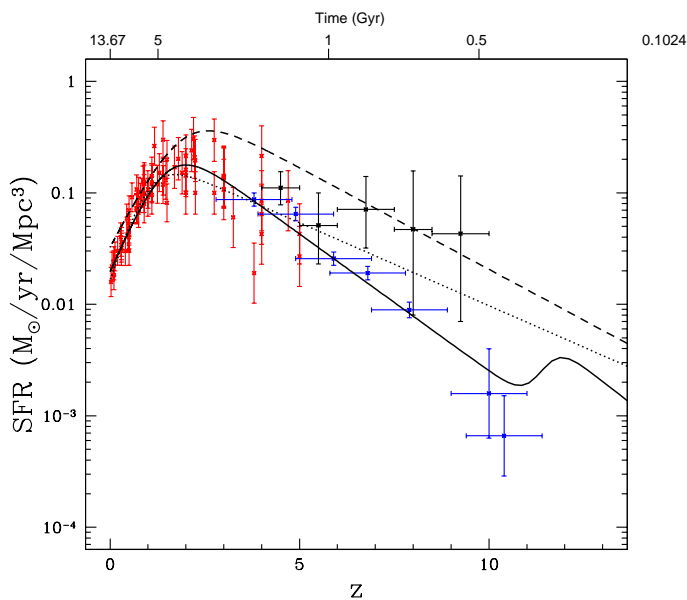, width=\linewidth}
\end{center}
\caption{{\bf Cosmic SFR as a function of redshift.} Three SFR modes considered in the paper,  SFR1 (solid line), SFR2 (dotted line) and SFR3 (dashed  line). 
Observations are taken from \citet{Behroozi13} (red points), \citet{Bouwens14, O14} (and references therein) (blue points) and \citet{Kistler13} (black points).
\label{fig:sfr}
}
\end{figure}

The cosmic SFR is a key ingredient to model the rate of stellar explosive events, such as CCSN and NSM. The redshift evolution of the SFR density is now rather well constrained by many observations.
Recent data from the Hubble Ultra Deep Field have significantly
extended the range of redshift for its determination,
from 
 redshift $z =4$ up to 10 \citep{B07,B08,B11, Bouwens14, O12, O13, O14}. 
Even more recent observations of high $z$ galaxies and  GRBs
 tend to favor a large amount of still unseen SFR at $z>9$ \citep{Kistler13, wang13}.
This last study suggests that SFR density may only decline out at $z=11$ and 
 that GRBs may be useful in exploring the unseen faint dwarf galaxies at high redshift. \citet{Behroozi13} re-analyzed the average star formation histories from $z=0$ to 8 and obtained consistent results with observed galaxy stellar mass functions, specific star formation rates and SFR. 
Moreover, high-redshift galaxy evolution has also been predicted by  \citet{Behroozi14} who, including recent  
constraints
\citep{Behroozi13, O13, Kistler13, Trenti13}, probed the unobserved $z>8$ galaxy populations. The resulting observationally constrained SFR is illustrated in Fig.~\ref{fig:sfr} as a function of the redshift.
The corresponding evolution of the cosmic SFR density as a function of redshift can be parametrized by the following form proposed by \citet{sp03},
\begin{equation}
\psi(z) = \nu\frac{a\exp[b\,(z-z_m)]}{a-b+b\exp[a\,(z-z_m)]}\, ,
\label{eq:SFRparam}
\end{equation}
where $\nu$ (in $M_\odot$/yr/Mpc$^{3}$) and  $z_m$ correspond to the
 astration rate and the redshift
at the SFR maximum, respectively, while $b$ and $b-a$ 
fix
 the SFR slope at low and high redshifts, respectively. 
 
Due to the fundamental importance of the SFR 
in predicting the stellar explosive event rates, and more generally the cosmic chemical evolution,
we consider
in the present paper
three 
SFR modes (see Fig.~\ref{fig:sfr}),
which
coincide at low redshift, where they reproduce the available data, and differ at high redshift, taking into account the current uncertainties:
\begin{itemize}
\item 
SFR1 includes 
a standard mode of
 population (Pop) II/I stars formation between $0.1$ 
  and $100$ M$_\odot$, and additionally a Pop III stellar mode at high redshift between 36  
   and 100 M$_\odot$.
 The corresponding parameters for each mode in Eq.~(\ref{eq:SFRparam}) are, respectively,
$\nu=  0.18$  and $0.0025$, 
$z_m = 2.0$ and $12.0$,  
$a=2.37$  and $4.0$, 
and
$b =1.80$  and $3.36$.

\item 
SFR2 includes 
a unique mode of star formation
 between 0.1 M$_\odot$   and 100 M$_\odot$, with the following parameters:
$\nu =  0.15$, $z_m = 1.7$,  $a =2.8$, $b  =2.45$;

\item SFR3 also includes a unique mode of star formation
 between 0.1 M$_\odot$   and 100 M$_\odot$, with the following parameters:
 $\nu =  0.36$, $z_m = 2.6$,  $a =1.92$, $b  =1.5$.

\end{itemize}
For each mode, the IMF slope is set to the Salpeter value, i.e. $x=1.35$.

Star formation is assumed to start at the initial time of $t_0=100$~Myr 
corresponding to a redshift of $z=30$.
We 
adopt the following cosmological parameters, $\Omega_m = 0.27, \Omega_{\Lambda} = 0.73$ and $H_0= 71$ km/s/Mpc (h = 0.71) and a primordial power spectrum with a power-law index $n = 1$. 
The age $t$ of the Universe is then related to the redshift by
\begin{equation}
\frac{\mathrm{d}t}{\mathrm{d}z} = \frac{9.78\, h^{-1} \, \mathrm{Gyr}}{(1+z)\left[\Omega_{\rm \Lambda}+\Omega_{\rm m}(1+z)^{3}\right]^{0.5}}\, .
\label{eq1}
\end{equation}

As seen in Fig.~\ref{fig:sfr}, the SFR2 mode is an intermediate case, whereas SFR1 and SFR3 are respectively close to the lower and higher values of the current measurements at high redshift. We checked that each of these three scenarios is compatible with a large set of independent observational constraints discussed in Sect.~\ref{sec:evolution}, especially the Thomson optical depth to the cosmological micro-wave background (CMB) and the cosmic evolution of [Fe/H]. This led to the inclusion of the additional Pop III component at high redshift in the SFR1 case, the standard mode alone being too weak in this case to fulfill the reionization constraint (see Sect.~\ref{sect_wmap}).

\subsection{The NSM rate}
\label{sect_rates}

\subsubsection{The CCSN rate and the NS birth rate}

\begin{figure}
\begin{center}
\epsfig{file=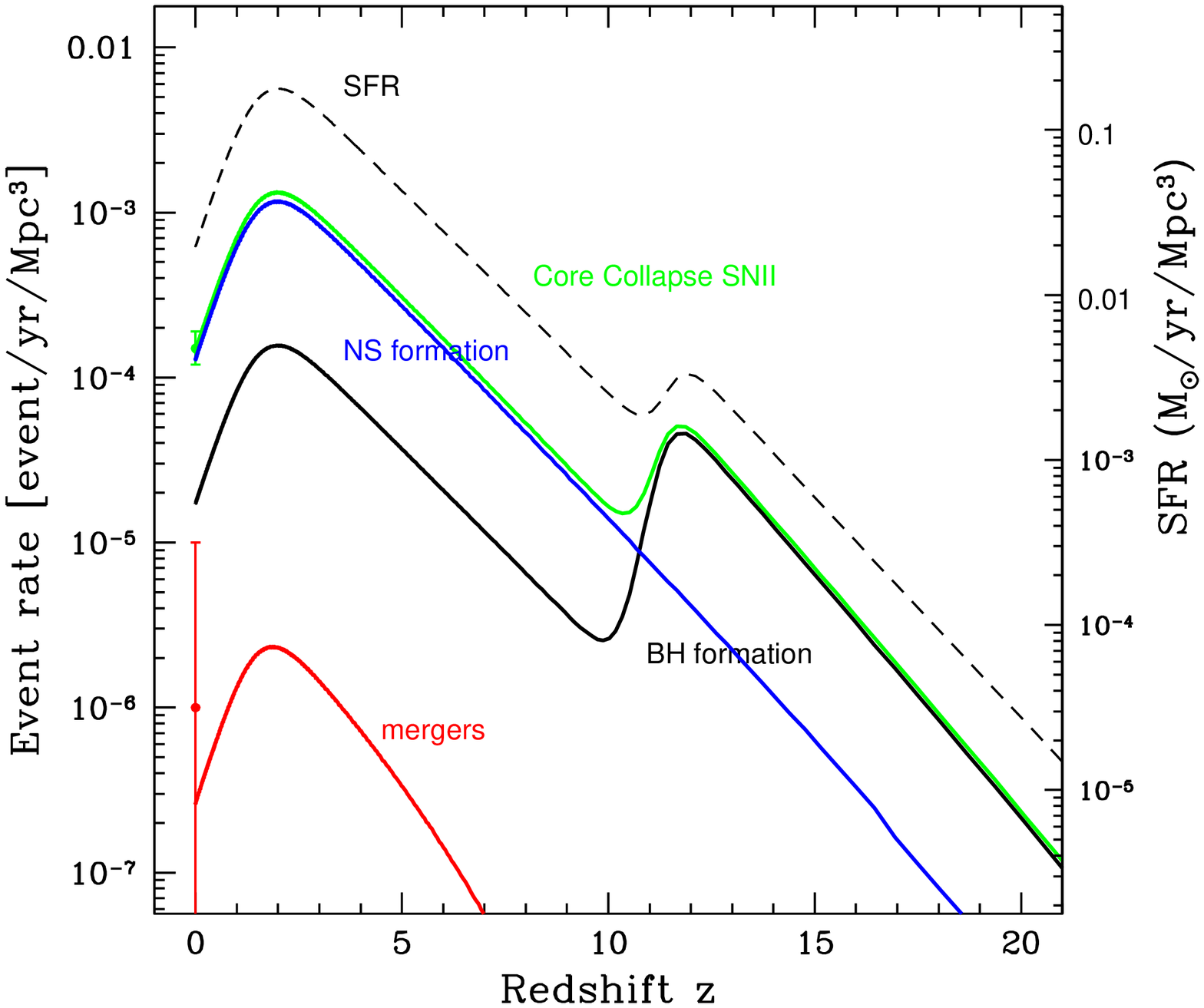, height=3in}
\vskip -1 cm
\epsfig{file=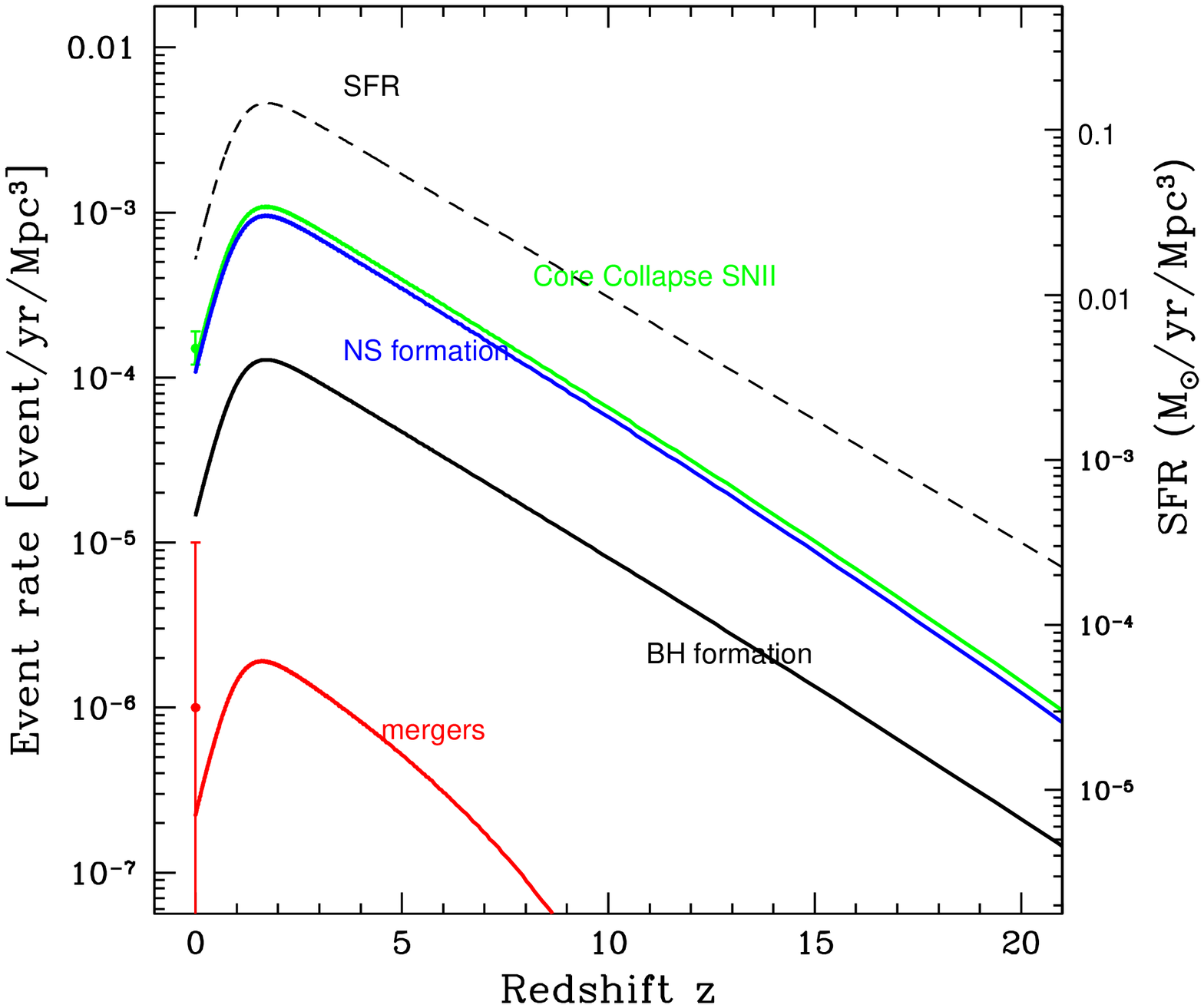, height=3in}
\end{center}
\caption{NS and BH birth rates as well as the CCSN and merger rates as a function of the redshift for SFR1 (upper panel) and SFR2 (lower panel).
Green and red points represent the local CCSN rate \citep{mattila12} and  the total local merger rate \citep{abadie10}, respectively. 
The SFR evolution is also plotted for comparison.
\label{fig:explosions}
}
\end{figure}

Once the evolution of cosmic star formation rate is known, the CCSN rate can be directly deduced if the mass-range of the progenitors is known. We use the predictions from \citet{ww95} and assume that the remnant of stars with mass $8 < M/M_\odot < 30$ is a NS, and that more massive stars produce a BH.
%
 This leads to the following NS birth rate at time $t$ (related to redshift $z$ by Eq.~\ref{eq1}):
\begin{equation}
\dot{R}_\mathrm{NS}(t)
=\int \mathrm{d}m\,  \Phi(m) \Psi(t_*) \, \Xi_\mathrm{NS}\left(  m , Z(t_*)  \right)\, ,
\end{equation}
where $\Phi(m)$ is the initial mass function (IMF) integrated over the whole stellar mass range, $\Psi(t)$ the SFR, $t_*$ the formation time of the progenitor, such that $t_*+\tau\left(m,Z(t_*)\right)=t$, $Z(t_*)$ the metallicity at the epoch of formation of the progenitor, $\tau\left(m,Z(t_*)\right)$ the lifetime of a star formed with mass $m$ and metallicity $Z(t_*)$, and  $\Xi_\mathrm{NS}\left(m,Z(t_*)\right)$ is $1$ if the stellar remnant is a NS, $0$ otherwise. The lifetimes of stars come from \citep{mm89} and the stellar metallicity-dependent remnants are taken from \citep{ww95}. The BH birth rate is given by a similar equation, using now $\Xi_\mathrm{BH}\left(m,Z(t_*)\right)=1$ if the remnant is a BH, $0$ otherwise, and the CCSN rate is the sum of these two rates.
The corresponding evolution of these three rates deduced on the basis of SFR1 and SFR2 is plotted in Fig.~\ref{fig:explosions}. On the observational point of view, the local CCSN rate is well determined, thanks to large surveys dedicated to the search of supernovae \citep{mattila12}. As can be seen in Fig.~\ref{fig:explosions}, the agreement with the prediction deduced from the cosmic SFR is good.

\subsubsection{The NS merger rate}

In contrast to the to the CCSN rate, the astrophysical estimate of compact-binary coalescence rates depends on a number of assumptions and unknown model parameters and are still rather uncertain \citep{abadie10}. Among the various estimates, the rate predictions for NS-NS binary systems is the most reliable one since they are based on extrapolations from observed binary pulsars in our Galaxy. These estimates lead to a coalescence rate of about 100 Myr$^{-1}$ per Galaxy, although this rate could plausibly range from 0.1 Myr$^{-1}$ to 1000 Myr$^{-1}$ \citep{kim03,kalogera04,men14}. 
Due to the lack of observational data, predictions of the
frequency of NS-BH mergers remains even more uncertain and rely exclusively on theoretical studies
\citep{tutukov93,voss03, oshaugh08, dom12, dom13,men14,postnov14}. These
population synthesis models estimate galactic merger rates between
$2\times 10^{-9}$ and $10^{-5}$ per
year. The range reflects the challenge in
comprehensively modelling the formation and evolution of stellar
binaries and their remnants. \citet{baus14} determined an independent upper limit on the merger rate
of NS-BH binaries by comparing the predicted r-process nucleosynthesis
yields of such systems with the observed galactic amount of r-process
material. A strict upper limit of the average NS-BH merger rate of about 6$\times 10^{-5}$
 per year
was found. \citet{dom13} calculated the cosmological merger rates of NS-NS, NS-BH, BH-BH systems as a function of redshift assuming different SFR histories. While in most cases NS-NS systems dominate the merger rates in the local Universe, BH-BH mergers are found to dominate at high redshift.

In the present study, the merger rate is determined assuming that 
the formation rate of NS-NS/BH binary systems is a fraction $\alpha$ of the NS birth rate $\dot{R}_\mathrm{NS}$ and that these binary systems merge after a delay $\Delta t_\mathrm{NSM}$, the value of which is discussed below. This leads to the following merger rate at time $t$:
\begin{equation}
\dot{R}_\mathrm{NSM}(t)
=\alpha\dot{R}_\mathrm{NS}\left(t-\Delta t_\mathrm{NSM}\right)\, .
\label{eq:Rnsm}
\end{equation}
In our model, both $\alpha$ and $\Delta t_\mathrm{NSM}$ and taken as free parameters.
The predicted evolution of the merger rate is plotted in Fig.~\ref{fig:explosions} for SFR1 and SFR2, assuming $\alpha=0.002$ and $\Delta t_\mathrm{NSM}=0.2$ Gyr. The independent estimate of the local rate taken from the compilation by \citet{abadie10} is indicated for comparison.


\subsubsection{The coalescence timescale}
\label{sec:mergerdelay}

The  coalescence timescale $\Delta t_\mathrm{NSM}$ is defined as the delay between
the formation of the binary system of two NS or a NS and a BH (note that it does not include the lifetime of the two star progenitors of the compact objects). It is a key ingredient as it leads to a different cosmic evolution of NSM compared to CCSN, for which there is no additional delay after the formation of a NS/BH.

The angular momentum loss due to the emission of gravitational waves governs the coalescence timescale. Its value strongly depends on the properties of the binary system, and in particular on the initial separation $a$ (or orbital period $T$) since $\Delta t_\mathrm{NSM}\propto a^{4}$ (resp. $\Delta t_\mathrm{NSM}\propto T^{8/3}$) (\citealt{peters63}, see also \citealt{kalogera01,hughes09}). This dependence leads to an expected broad range of values for $\Delta t_\mathrm{NSM}$, with however large uncertainties related to the distribution of possible initial orbital parameters in NS/NS and NS/BH binaries \citep[see][for a recent review]{hughes09}. 
For a binary system of two 1.4 $M_\odot$ NS, the coalescence timescale is $\Delta t_\mathrm{NSM}=64$~Myr for $a=10^{-2}$~AU (resp. $T=5.2$~h).
Only seven NS/NS binary systems are known with measured masses \citep{lorimer05,lorimer08}. The current orbital parameters and the measured masses allow us to estimate the remaining time before the merger. As at least one of the two NS is observed as a pulsar, the measurement of the pulsar period and its time derivative can provide an estimate of the current age of the binary system. The sum of these two timescales gives an estimate of $\Delta t_\mathrm{NSM}$. 
Using the values listed in \citet{lorimer08}, 
coalescence timescales between 100 and 400 Myr are found in four cases, and values larger than 1 Gyr in the three other cases. The double pulsar PSR J037-3039 corresponds to the lowest value, $\Delta t_\mathrm{NSM}\simeq 180$ Myr. The famous binary pulsar PSR B1913+16, which has been studied in details for years as a test of general relativity \citep{hulse75,weisberg10} is characterised by a coalescence timescale $\Delta t_\mathrm{NSM}\simeq 420$ Myr.

\begin{figure}
\begin{center}
\epsfig{file=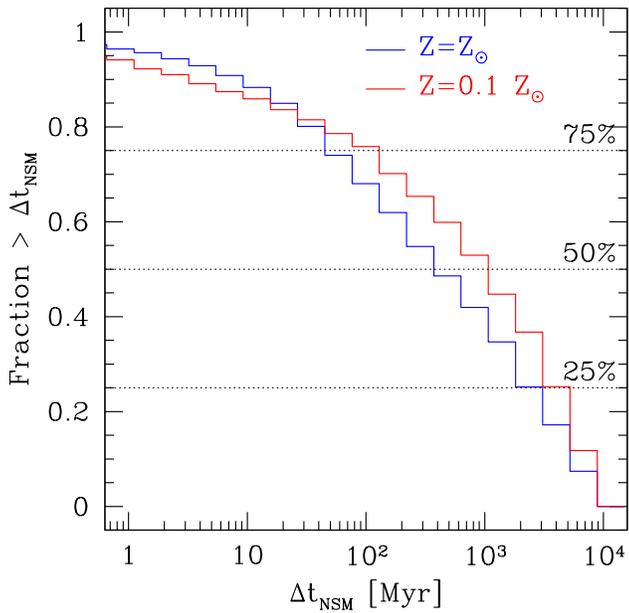, width=\linewidth}
\end{center}
\caption{{\bf Distribution of the coalescence timescale $\Delta t_\mathrm{NSM}$ for NS/NS binary systems.}
The coalescence timescale $\Delta t_\mathrm{NSM}$ is defined as the delay between the formation of the two NS and the final merger.
The plot shows the fraction of mergers with a coalescence timescale above a given value in Myr, for two different metallicities.
This distribution has been obtained from an updated version of the population synthesis model described in \citet{bel02}. It covers a broad range of values with a distribution falling approximately like $p(\Delta t_\mathrm{NSM})\propto \Delta t_\mathrm{NSM}^{-1}$.
}

\label{fig:delay}
\end{figure}

 The distribution of the coalescence timescale can be estimated using a population synthesis code for binary systems, which can be calibrated with observed systems. The result is plotted for two different metallicities in Fig.~\ref{fig:delay}. It has been obtained using the  {\tt StarTrack} code described in \citet{bel02,Belczynski08}, where standard stellar evolution models are implemented with the most updated physics regarding the relevant processes for the evolution of binary systems. Fig.~\ref{fig:delay} has been produced using the results of a recently updated version of the code
(Mink \& Belczynski in preparation), including new prescriptions for stellar wind mass loss rates \citep{Vink1, Vink2}, for supernova models \citep{Fryer12}, for the common envelop phase \citep{dom12}, adding electron capture supernovae for NS formation \citep{Belczynski08}, and taking into account recent observational constraints on massive stars in binary systems \citep{Sana12}. 
Additionally, the evolutionary
model included in the code does not allow for the progenitors to undergo a common-envelope phase  while one of the massive binary stars is evolving through the Hertzsprung gap \citep[see][for more details]{bel07}.
This corresponds to a  
conservative assumption and additional NS-NS mergers may form  if this assumption is relaxed.
Compared to the results of  \citet{bel02}, this updated calculation predicts a fraction of undetected galactic merging NS-NS systems significantly smaller, now at the level of $\sim 10$~\%.
 As illustrated in Fig.~\ref{fig:delay},

the distribution of coalescence timescales predicted by such a detailed calculation shows a broad range of values: at solar metallicity $Z=Z_\odot$ (resp. at metallicity $Z=0.1 Z_\odot$), it is found that
29\% (resp. 28\%) of systems have $\Delta t_\mathrm{NSM}\le 100$~Myr, 
13\% ( resp. 16\%) have $100 < \Delta t_\mathrm{NSM}\le 300$~Myr, 
15\% (resp. 14\%) have $0.3 < \Delta t_\mathrm{NSM}\le 1$~Gyr, 
17\% (resp. 16\%) have $1 < \Delta t_\mathrm{NSM}\le 3$~Gyr, 
and  26\% (resp. 26\%) have $\Delta t_\mathrm{NSM} > 3$~Gyr.
The impact of the metallicity is small.
 It appears clearly that a significant fraction of systems has a coalescence timescale of a few 100 Myr or less. These systems may dominate the early evolution of the NS merger, and therefore the evolution of Eu at redshift discussed in Sect.~\ref{sect_res}.

\section{Cosmic chemical evolution}
\label{sect_cem}
\subsection{Model description}

In addition of the rates of explosive events (CCSN, NSM) discussed in the previous sections, the predicted chemical evolution can also be deduced from the cosmic evolution  of the SFR. We use
the model developed by \citet{daigne06, rollinde09, vangioni14}, which is based on a hierarchical model for structure formation \citep{ps74,  sheth:99, jenkins:01, wyithe:03}.  We assume that the minimum mass of dark matter haloes for star-forming structures is $10^{7}~M_\odot$.

Finally, for a given evolution of the cosmic SFR density, as discussed in \S~\ref{sec:sfr}, the model follows the evolution of baryons in the Universe in terms of two main reservoirs. The first  is associated with collapsed structures and is divided in two sub-reservoirs: the ISM gas (mass $M_\mathrm{ISM}(t)$) and the stars and their remnants (mass $M_{*}(t)$). The second reservoir, the intergalactic medium (IGM, mass $M_\mathrm{IGM}(t)$), corresponds to the medium in between the collapsed structures. The evolution of the baryonic mass of these reservoirs is governed by a set of differential equations:
\begin{equation}
\frac{dM_\mathrm{IGM}}{dt}=-\frac{dM_\mathrm{struct}}{dt}=-a_\mathrm{b}(t)+o(t),
\end{equation}

\begin{equation}
\frac{dM_\mathrm{*}}{dt}=\Psi(t)-e(t)\ \mathrm{and}\ \frac{dM_\mathrm{ISM}}{dt}=\frac{dM_\mathrm{struct}}{dt}-\frac{dM_\mathrm{*}}{dt}\ .
\end{equation}

\noindent In addition, we have $M_\mathrm{ISM}(t)+M_\mathrm{*}(t)=M_\mathrm{struct}(t)$,
corresponding to the total baryonic mass of the structures, and
 $M_\mathrm{IGM}(t)+M_\mathrm{struct}(t)=\mathrm{constant}$ (total baryonic mass of the Universe).

 As can be seen, these equations are controlled by four rates which represent four fundamental processes: 
 the formation of stars  through the transfer of baryons from the ISM, $\Psi(t)$, which is directly fixed by the assumption SFR1, 2 or 3 discussed in \S~\ref{sec:sfr}; 
the formation of structures through the accretion of baryons from the IGM, $a_\mathrm{b}(t)$;
the ejection of enriched gas by stars, $e(t)$ and the outflow of baryons from the structures into the IGM, $o(t)$.

In the model, we track the chemical composition of the ISM and the IGM as a function of time or redshift. The differential equations governing the evolution of the mass fraction $X_{i}^\mathrm{ISM}$ ($X_{i}^\mathrm{IGM}$) of element $i$ in the ISM (IGM) are given by equations (6) and (7) in section 2 in \cite{daigne04}.

The baryon accretion rate, $a_\mathrm{b}(t)$ is computed in the framework of the hierarchical scenario of structure formation and the baryon outflow rate, $o(t)$, from the structures includes a redshift-dependent efficiency, which accounts for the increasing escape velocity of the structures as the galaxy assembly is in progress.

We use the framework of the hierarchical scenario where small structures are formed first. At redshift $z$, the comoving density of dark matter halos in the mass range $[M,M+dM]$ is $f_\mathrm{PS}(M,z)dM$, with 
\begin{equation}
\int_{0}^{\infty} dM\ M f_\mathrm{PS}(M,z) = \rho_\mathrm{DM}\ ,
\end{equation}
where $\rho_\mathrm{DM}$ is the comoving dark matter density. The distribution function of halos $f_\mathrm{PS}(M,z)$ is computed using the method described in \cite{jenkins:01}. We adopt a rms amplitude $\sigma_{8}=0.9$ for mass density fluctuations in a sphere of radius $8\ h^{-1}\ \mathrm{Mpc}$.
We assume that the baryon distribution traces the dark matter distribution without any bias so that the density of baryons is just proportional to the density of dark matter by a factor $\Omega_{b}/\left(\Omega_{m}-\Omega_{b}\right)$. 

The fraction of baryons at redshift $z$ which are in such structures is given by
\begin{equation}
f_\mathrm{b,struct}(z) = \frac{\int_{M_\mathrm{min}}^{\infty} dM\ M f_\mathrm{PS}(M,z)}{\int_{0}^{\infty} dM\ M f_\mathrm{PS}(M,z)}\ .
\label{eq:fb}
\end{equation}
Therefore, the mass flux $a_\mathrm{b}$ can be estimated by
\begin{eqnarray}
a_\mathrm{b}(t) & = & \Omega_\mathrm{b}\left(\frac{3H_{0}^{2}}{8\pi G}\right)\ \left(\frac{dt}{dz}\right)^{-1}\ \left|\frac{d f_\mathrm{b,struct}}{dz}\right|\, .
\end{eqnarray}

In addition to the NS and BH birth rates, the CCSN rate, and the NS-NS and NS-BH merger rates discussed in \S~\ref{sect_rates} (and of special interest in the present study), several other
 quantities are
  followed as a function of the redshift $z$, namely
    the WD birth rate, 
the
type Ia supernova rate, 
  the abundances of various chemical elements, including Fe and Eu, in the ISM and the IGM,
  the ionizing flux from stars, the ionization state of the IGM and the Thomson optical depth of the CMB. 
 A detailed description of this cosmic evolution model 
  can be found in \citet{daigne06,rollinde09,vangioni14}.

\subsection{Observational constraints on the chemical evolution model}
\label{sec:evolution}

We tested that each cosmic evolution scenario discussed in this paper fulfills two important observational constraints,
namely the observed cosmic evolution of  [Fe/H]  (corresponds to the logarithm of the abundance of iron normalized to the solar one) and the measured Thomson optical depth of the CMB.
Following the detailed analysis presented in \citet{vangioni14} for different possible SFR using the same cosmic evolution model, this leads to the choice of the three SFRs presented in \S~\ref{sec:sfr}. The SFR1 and SFR2 cases are directly taken from their study. 

SFR1
corresponds to a standard SFR with an additional Pop III star component at high redshift,
mainly motivated by the fact that 
the ionizing flux at high redshift produced by the standard mode only
 is not large enough to reproduce the Thomson optical depth of the CMB (see  \S~\ref{sect_wmap}).

SFR2 
 is a SFR enhanced at high $z$. 
 The third mode considered here, SFR3, is chosen as the upper limit of the SFR at high redshift obtained from the constraint of the Thomson optical depth of the CMB (see Fig.~\ref{fig:ionisation}).
 As shown in Fig.~\ref{fig:sfr}, SFR1 fits
 the observational constraints from \citet{Behroozi13}; \citet{Bouwens14} and \citet{O14}, whereas SFR2 and SFR3  take into account the studies of \citep{Trenti13} and \citet{Kistler13}, respectively. 
%
%
%


\subsubsection{Reionization}
\label{sect_wmap}

\begin{figure}
\begin{center}
\epsfig{file=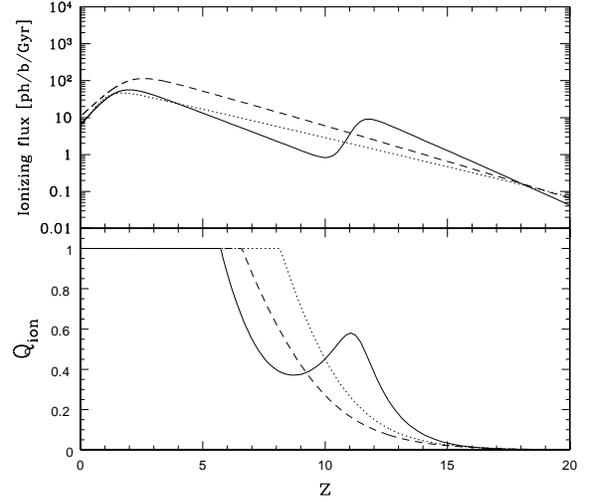, height=3in}
\vspace*{-1cm}
\epsfig{file=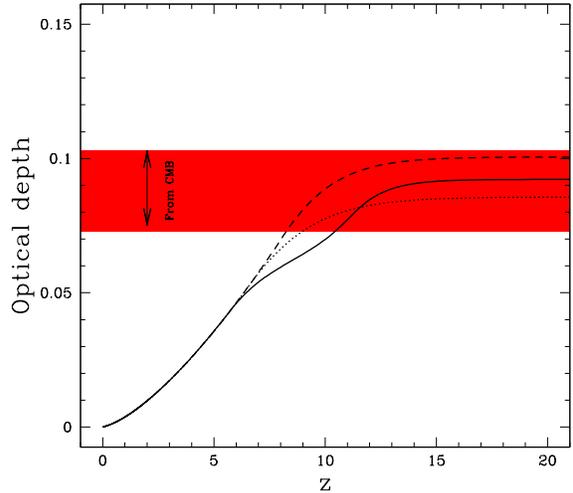, height=3in}

\end{center}

\
\caption{{\bf Reionisation.} Evolution
as a function of redshift of
the ionizing flux (top panel), volume filling fraction $Q_{{\rm ion}}$ (middle panel) 
and Thomson optical depth (lower panel) for the three  
SFR modes (conventions are the same as
 in Fig.~\ref{fig:sfr}). 
WMAP9 results for the optical depth at high redshift \citep{wmap} are indicated by a red strip in the lower panel.
}
\label{fig:ionisation}
\end{figure}

Concerning the ionization history, the 
evolution of the volume filling fraction $Q_\mathrm{ion}(z)$ of ionized regions at redshift $z$ is given by
\beq
\frac{\mbox{d}Q_{{\rm ion}}(z)}{\mbox{d}z} = \frac{1}{n_{\rm
 b}}\frac{\mbox{d}n_{{\rm ion}}(z)}{\mbox{d}z}-\alpha_{{\rm B}}n_{{\rm b}}C(z)
   Q_{{\rm ion}}^{2}(z)\left(1+z\right)^{3}\left|\frac{\mbox{d}t}{\mbox{d}z}\right|\mbox{\ ,}
\eeq
where $n_{\rm b}$ is the comoving density in baryons, $n_{{\rm ion}}(z)$ the comoving density
of ionizing photons,  $\alpha_{{\rm B}}$ the recombination coefficient,  and $C(z)$ the
clumping factor. This last factor is  taken from \citet{greif06} and  varies from a value of 2 at $z\leq20$ to a constant value of 10 for 
$z<6$. The escape fraction,  $f_{\rm esc}$, is set to 0.2 for each of our SFR modes. The number of ionizing photons 
emitted by massive stars is calculated using the tables given in \citep{schaerer02}. Finally, the Thomson optical
depth 
between redshifts $z'=0$ and $z$
is computed following \citep{greif06}, i.e.
\begin{equation}
\tau =c\,\sigma_{{\rm T}}\,n_{\rm b} \int_{0}^{z}\mathrm{d}z'\,Q_{{\rm ion}}(z')\left(1+z'\right)^{3}\left|\frac{\mbox{d}t}{\mbox{d}z}(z')\right|\mbox{\ ,}
\end{equation}
where 
$\sigma_{{\rm T}}$ is the Thomson scattering
 cross section. 
 Fig.~\ref{fig:ionisation} shows the 
 evolution of the volume filling fraction $Q_{{\rm ion}}$ and the ionizing flux 
   for the three SFR histories, and the resulting Thomson optical depth,
the value of which is found to agree well with the measurement from CMB observations by 
WMAP9  \citep{wmap} at high redshift.
As mentioned above, the SFR1 scenario without the Pop III mode at high redshift would not fulfill this constraint
  \citep{vangioni14} and SFR3 is adjusted to reproduce the upper limit on the optical depth.
Note that the last results  from the \citet{planck14, planck15} give 
a revised Thomson optical depth $\tau = 0.079 \pm 0.017$. This value is lower 
than the WMAP9 one ( $\tau = 0.089 \pm 0.014$). Consequently, in this context, the contribution of Pop III stars in the SFR1 model should be lowered.

\subsubsection{Cosmic chemical evolution: iron}
\label{sect_in}
Our model includes the stellar lifetimes estimated by  \citet{mm89} for intermediate mass stars with $0.9< M/M_\odot < 8$ and by \citet{schaerer02} for more massive stars. 
Lifetimes of low mass stars  ($M/M_\odot < 0.9$), such as the old halo stars where abundances of Eu are measured at low metallicity, are long enough for the star to be still observed today, whatever the redshift of formation.
Stars formed at redshift $z$ inherit their initial chemical composition from the abundances in the ISM at the same redshift.
Thus, the observed abundances in stars reflect, in a somewhat complex way, the yields of all massive stars that have exploded earlier. The stellar yields adopted in our model depend on the stellar mass and metallicity. 
By default, we use  the tables of yields  from \citep{ww95}  for massive stars ($8< M/M_\odot<40$).
 An interpolation is made between the different metallicities ($Z=0$, $0.0001 Z_\odot$, $0.001 Z_\odot$, $0.1 Z_\odot$ and $Z_\odot$) and masses,  and tabulated values are extrapolated for stars beyond $40\,M_\odot$, which correspond to a small fraction of the population of massive stars 
 (i.e. $\left(40^{-x}-100^{-x}\right)/\left(8^{-x}-100^{-x}\right)\simeq 8\%$  for the Salpeter IMF of index $x = 1.35$).

\begin{figure}
\begin{center}

\epsfig{file=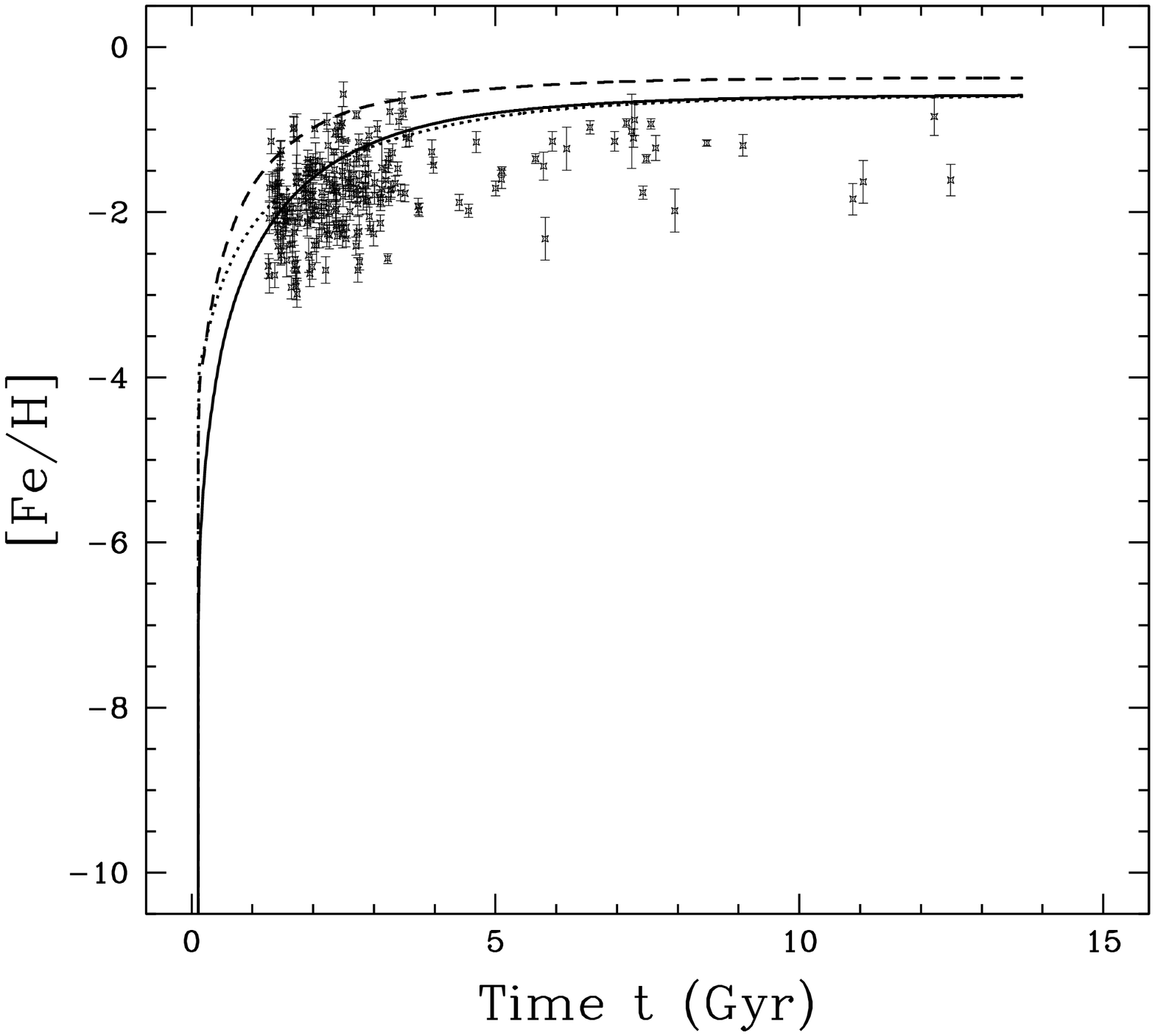, height=3in}
\vspace*{-0.5cm}
\epsfig{file=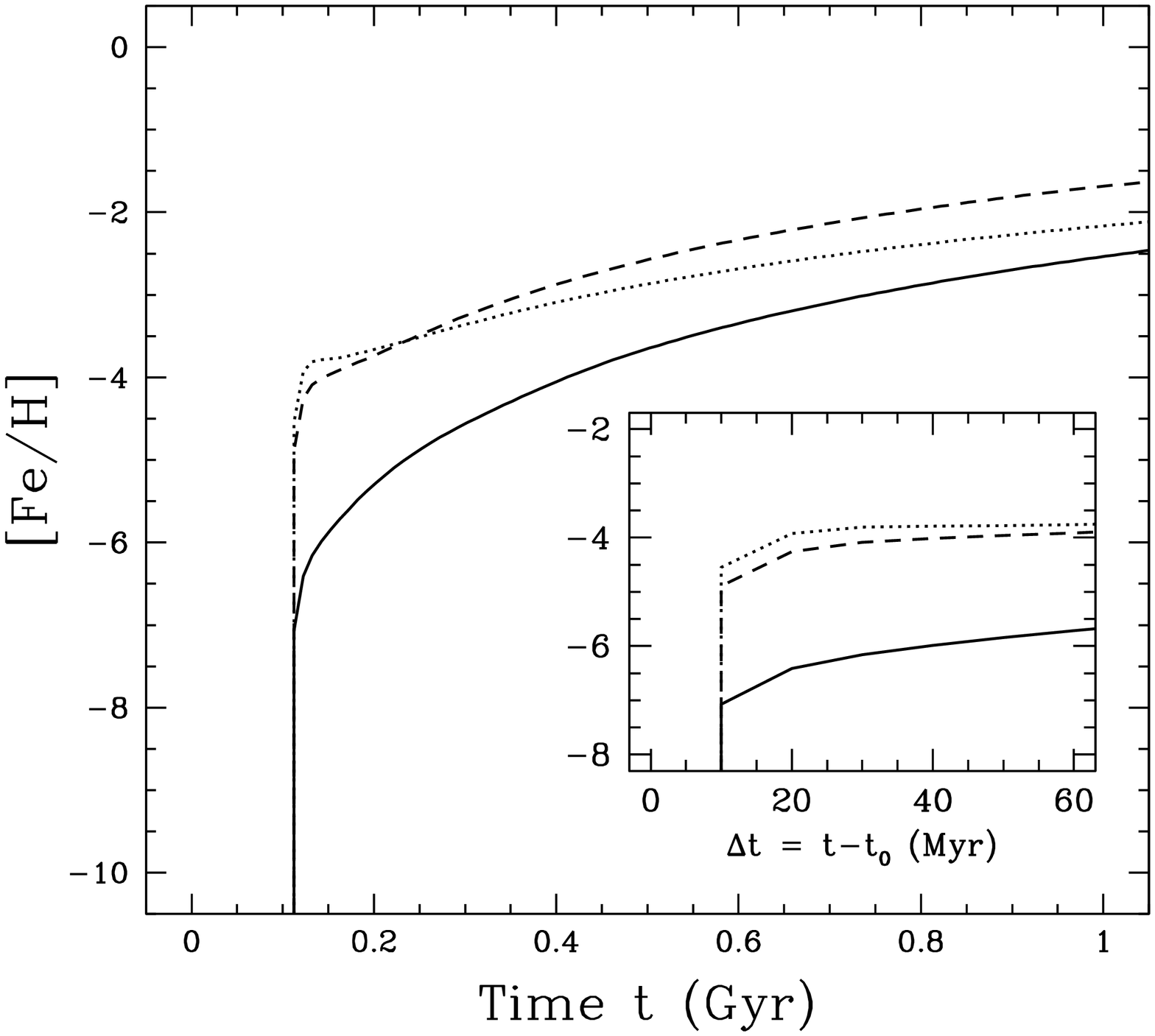, height=3in}
\vspace*{-0.5cm}
\epsfig{file=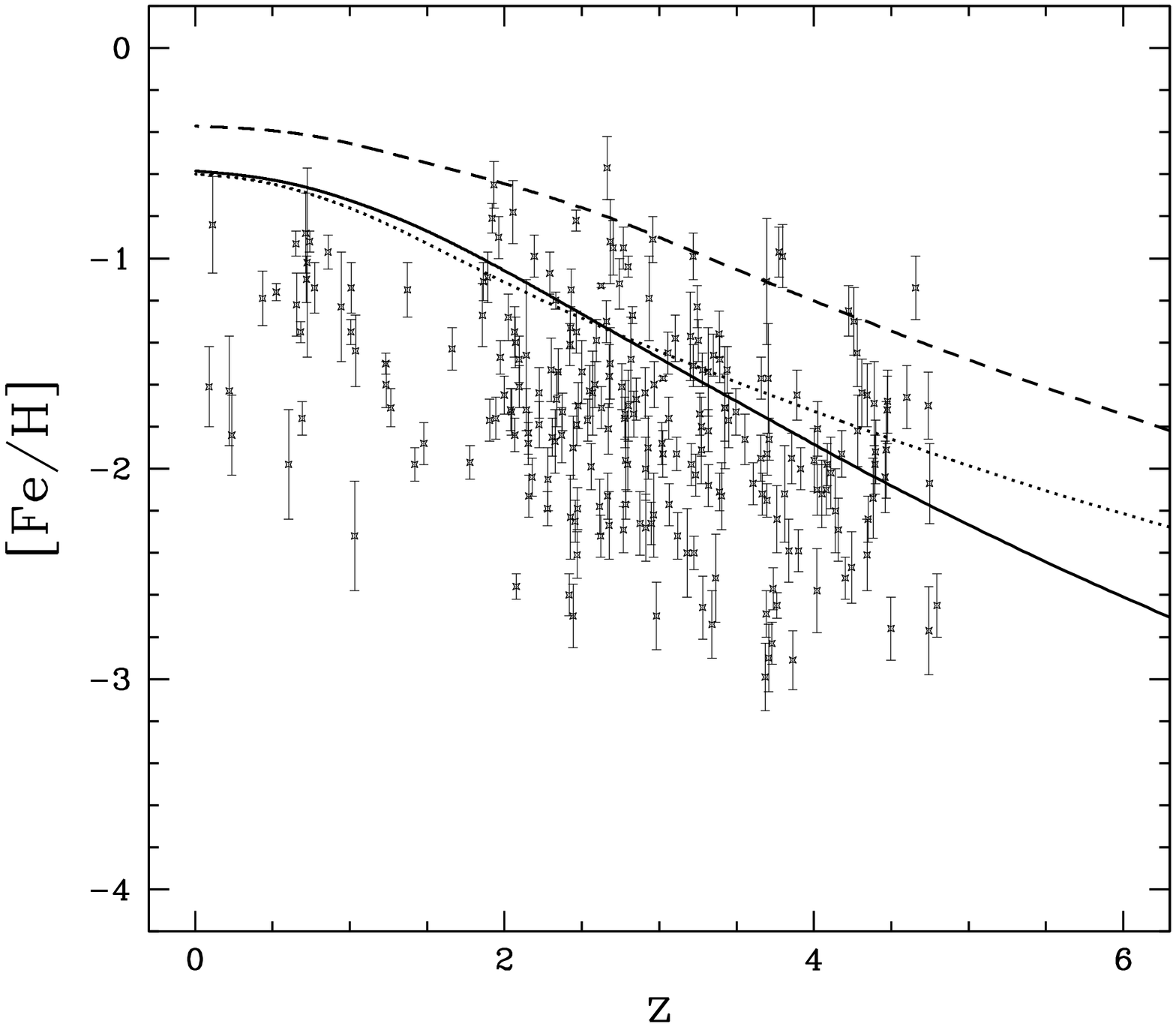, height=3in}

\end{center}
\
\caption{{\bf Iron evolution.} [Fe/H] evolution as a function of  time $t$ (upper panels) or redshift $z$ (lower panel) for the three SFR modes (lines are the same as defined in Fig.~\ref{fig:sfr}). 
The middle panel focusses on the early time evolution, with the insert emphasizing the three different time delays around the [Fe/H]=-4 iron enrichment .  Data correspond to DLAs observations as a function of redshift \citep{raf12}. Eq.~(\ref{eq1}) is used to convert $z$ into $t$ when including data in the time evolution panels.
}
\label{fig:iron}
\end{figure}

Fig.~\ref{fig:iron}  shows the [Fe/H] evolution as a function of time and redshift for the three SFR modes. Data correspond to chemical abundance determinations in about 250 damped Ly$\alpha$ (DLA) systems  \citep[][ and references therein]{raf12} as a function of the redshift.  In their paper, they present 47 DLA systems at $z>4$ observed with the Echellette Spectrograph and Imager and the High Resolution Echelle Spectrometer on the Keck telescopes \citep[for more details see][]{vangioni14}. We use Eq.~(\ref{eq1}) to associate a time with the observed redshift for each observed DLA system.
As illustrated in Fig.~\ref{fig:iron}, the cosmic evolution of the iron abundance is reproduced by each of the three scenarios,
 passing through the upper limit of the points. However, note that these iron abundances are measured in the gas phase which should be considered as lower limits due to depletion into grains in the ISM.
As seen in Fig.~\ref{fig:iron} (middle panel), the early iron evolution
depends strongly on
 the shape of the SFR at high $z$. In particular, it can be seen that 
 a cosmic [${\rm Fe/H]}=-4$ enrichment is achieved at the very beginning, 
$\Delta t=t-t_0\simeq 0.015$~Gyr for SFR2, 0.05~Gyr for SFR3, and  0.4~Gyr for  SFR1. This time delay in the early Fe enrichment can have a drastic impact on the interpretation of the early cosmic chemical evolution with respect to observations, as discussed in Sect.~\ref{sect_res}.

\subsubsection{Cosmic chemical evolution: other elements}
\label{sect_in2}

\begin{figure}
\begin{center}
\begin{tabular}{c}
\epsfig{file=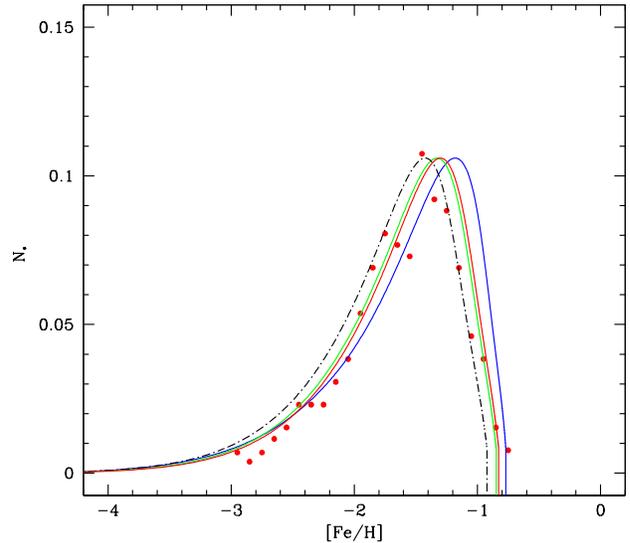, width=\linewidth}
\end{tabular}
\end{center}
\caption{{\bf Local metallicity distribution function}, i.e.
  normalized number of stars as a function of the iron abundance [Fe/H].  The MDF is calculated with SFR1 mode and the red circles to the observed MDF deduced by \citep{an13} using SDSS photometry. We present four curves obtained with different Fe yields reflecting  uncertainties mainly related to the CCSN mass cut. Blue curve : standard yields from \citep{ww95} with yields varying with metallicity, green curve : same after dividing these yields by a factor 2, dot-dash black curve : same using only yields calculated from solar metallicity stars only, red curve : same when using the iron yields determined by \citep{Koba06}.
}
\label{fig:hist}
\end{figure}

As detailed  in \cite{vangioni14}, the present cosmic evolutionary model has been used to predict the ``standard'' chemical evolution of light elements, such as C, N, O or Mg and to test its predictive power with respect to available observations \citep{suda08, suda11}. The sensitivity to different SFR has also been explored. 
\begin{itemize}
\item \textit{Carbon, oxygen.}
It was found that the [C/H] evolution as a function of the iron abundance is essentially independent of the adopted SFR model
 for [Fe/H] $> -2$ or $z < 4$.  At higher $z$ (lower metallicity), the SFR including a Pop III mode (like SFR1) was shown to be in better agreement with the observation of ultra-metal-poor  carbon-enhanced stars, which presents a significant dispersion. Similar conclusions were drawn for the oxygen evolution, where the Pop III contribution to the SFR1 model allows for a high production of O at low metallicities with a local peak in [O/H] at [Fe/H] $\approx -4.8$. Interestingly, this model can also explain the oxygen abundances in ultra-low-metallicity stars (see in particular Fig.~9 in \citealt{vangioni14}).

\item \textit{Magnesium.}
As in the oxygen case, the inclusion of a high-mass model in SFR1 at large redshift directly impacts the 
evolution of Mg in star forming structures.
The predicted Mg pattern in the SFR1 case is, however, not observed at low metallicity \citep[see Fig. 10 in][]{vangioni14}. However, new studies in high mass stars show that 
Mg evolution at low metallicity remains sensitive to the predicted nucleosynthesis in massive zero-metallicity stars. Such calculations have been recently revisited by \citet{hw10} who showed that in their models no Mg is produced in zero-metallicity stars more massive than 30 M$_\odot$. In this case, the early production of Mg in the three SFR cases discussed here is found to reproduce well the bulk of data, as shown in Figure 10b of \citet{vangioni14}.

\item \textit{Local stellar metallicity distribution function (MDF).}

We show in Fig.~\ref{fig:hist} the 
MDF
 derived by \citet{an13} using the SDSS photometric Catalogue together with our computed MDF based on the SFR1 mode (the other modes giving very similar results).
As found by \citet{an13}, the in-situ photometric metallicity distribution has a shape that matches that of the kinematically-selected local halo stars from \citet{no91}.  For this reason, we adopt their SDSS calibration catalog MDF as a reference since it presents the advantage of expressing the MDF directly  as a function of  [Fe/H].
The global shape of our computed  MDF is shown in Fig.~\ref{fig:hist} to be in good agreement with the observation. We point out here the impact of the choice of the iron yields and the related  
uncertainties. We present four curves reflecting  uncertainties mainly related to the CCSN mass cut. The blue curve corresponds to the standard yields from \citet{ww95} adopted in the rest of the paper, which are  
varying with metallicity, the green curve to the same yields divided by a factor 2, the dot-dash black curve to yields calculated from solar metallicity stars only, and the red curve to Fe yields determined by \citet{Koba06}.

\item{\textit{Stellar mass.}}
As shown by \citet{vangioni14}, the cosmic 
evolution of the stellar mass as a function of the redshift is also found to be in good agreement with cosmological observational photometric data (see their Fig. 12b), in particular for the SFR1 mode.
\end{itemize}

In conclusion, based on the previous study of  \citet{vangioni14} within the same cosmological context and on the basis of the same SFR cases, it can be concluded that the standard chemical evolution is rather well reproduced by the 
cosmic evolution of the 
star formation considered in the present study, 
and consequently that the present model is well suited to describe the averaged chemical history.
The evolution of Milky Way-like galaxies appears to be not far from the averaged history. As discussed recently in \citet{frebel15},
this opens the possibility of "near-field cosmology" based on stellar archeology, especially using the population of the oldest stars in the halo of our Galaxy and in dwarf galaxies. We will use this approach in the next Section to study the cosmic evolution of r-process elements.

\section{Cosmic evolution of r-process elements}
\label{sect_res}




As far as the galactic chemical evolution is concerned, CCSN remain promising sites, especially in view of their potential to significantly contribute to the galactic enrichment \citep{argast04}. However,  they remain handicapped by large uncertainties associated mainly with the still incompletely understood mechanism that is responsible for the  supernova explosion and the persistent difficulties to obtain suitable r-process conditions in self-consistent dynamical explosion and neutron-star cooling models \citep{janka12,hudepohl10,fisch10}.  In addition, nucleosynthesis predictions of the detailed composition of the ejected matter remain  difficult due to the remarkable sensitivity of r-process calculation to the still unknown initial properties of the ejecta.

In contrast, NS-NS or NS-BH mergers, which today are clearly favored from a nucleosynthesis point of view, have been 
claimed to be ruled out as the dominant r-process source by early studies
on the basis of inhomogeneous chemical evolution models \citep{argast04}, due to their low rates of occurrence which seem to be inconsistent with observations of low-metallicity r-process-rich stars. In addition, the significant amount of r-process material ejected by a single NSM leads to a large scatter in r-process enrichment at later times that does not seem to be confirmed by observations. 

Recent studies \citep{mat14,kom14,ces14,men14, tsu14a, tsu14b, shen14,voort14, enrico14} have reconsidered the chemical evolution of r-process elements in different evolutionary contexts, and reached rather different conclusions.
More specifically, \citet{mat14} 
explored the  Eu production in the Milky Way using a local chemical evolution model. Their observational  constraints at low metallicity come essentially from the observations of Eu in metal-poor stars by \citet{francois07}, without including the related error bars and upper limits. 
The relevance of the NSM scenario on the production of Eu has been studied by testing the effect of  {\it (i)} the  coalescence timescale of the binary system  {\it (ii)}  the Eu yield expected to be ejected from NSM and  {\it (iii)} the range of the initial mass of the NS progenitors. Similarly, the CCSN scenario has been explored by considering different possible Eu yields assuming a strong r-process taking place in CCSN. In this framework, NSM is found to be potentially a major r-process source if the coalescence timescale is of the order of 1 Myr and the ejected Eu yield of the order of  $3 \times 10^{-7}$ M$_\odot$ for a mass range of progenitors of NS ranging between 9  and 50 M$_\odot$. 
The scenario where both CCSN and NSM contribute to the Eu synthesis is also compatible with observations
provided NSM produce $2 \times10^{-7}$ M$_\odot$ of Eu per system and each CCSN with progenitors in the range of 20--50 M$_\odot$ produce around  $10^{-8} - 10^{-9}$ M$_\odot$ of Eu.

 In parallel, \citet{kom14} investigated the chemical enrichment of r-process elements using a hierarchical galaxy formation model. The CCSN scenario is found to reproduce the scatter of observed r abundances in low metallicity stars if about 10\% of CCSN in the low-mass end (i.e. for progenitor mass of the order of 10 M$_\odot$) is the dominant r-process source and the star formation efficiency amounts to about 0.1 per Gyr. For NSM to be the main r-process site, a coalescence timescale of about 10 Myr with an event rate about 100 times larger than currently observed in the Galaxy need to be considered. 
 
We note that the coalescence timescale in the studies by \citet{mat14} and \citet{kom14} is constrained to be surprisingly short compared to known systems in our Galaxy and theoretical predictions by stellar population models (see Sect.~\ref{sec:mergerdelay}).

 \citet{ces14} computed inhomogeneous chemical evolution models for the galactic halo, taking into account the contribution of electron-capture and magnetorotationally driven supernovae (including fast-rotating progenitors), but not NSM, to explain the Eu scatter in metal-poor stars.

 \citet{men14} also studied the temporal evolution of the galactic population of double NS binaries, mixed systems with a NS and BH component, and double BH binaries. They conclude that, except for the first 100~Myr of the evolutionary phase of the Galaxy, double compact star mergers may be the major production sites of r-process elements, and it is probable that the mixed NS-BH systems dominate over double NS binary mergers.

 Finally, \citet{shen14} and \citet{voort14} estimated the enrichment history of r-process elements in the Galaxy, as traced by the [Eu/Fe] ratio, using a high resolution cosmological zoom-in simulation. Unlike previous studies, it was found that the nucleosynthetic products from compact binary mergers can be incorporated into stars of very low metallicity and at early times, even with a minimum  time delay of 100 Myr and that compact binary mergers could be the dominant source of r-process nucleosynthesis.

In the present Section, we bring our own contribution to this difficult debate. We first summarize the available
Eu observations and then propose a detailed comparison with our predictions of the Eu evolution along the cosmic history, using the three possible cosmic SFR density discussed in \S~\ref{sec:sfr}.
%


\subsection{Eu: observations }
\label{sect_obs}

We have gathered  most of the recent observations on Eu abundance (Eu/H and [Eu/Fe] as a function of [Fe/H]) in metal poor stars \citep{francois07, honda04, barklem05, sim04, roed11, roed10, roed12, roed14a, roed14b, roed14c, ren12, wor13} together with the abundances found in dwarf spheroidal systems \citep{shertone01, shertone03, geisler05, cohen09, letarte10, Starkenburg13, McWilliam13}. It is interesting to note that the bulk of data coming from external dwarf galaxies is embedded in galactic ones.
The solar abundances are taken from \citet{lod03}.

The determination of Eu abundance suffers from the fact that only few transitions are visible in the  spectrum of metal poor stars. The strongest transitions are rather weak and it becomes particularly difficult to detect and estimate the abundance of this element at low metallicities. Typically, the strongest  Eu line becomes almost undetectable below [Fe/H] $\simeq -3.5$. The detection or the non-detection depends on the S/N ratio of the spectrum, on the presence or not of an overabundance of Eu  in a given star and finally on the temperature and gravity of the star. This situation clearly shows up in the diagrams where the number of true detection of Eu at very low metallicity are only a handful.  Moreover, due to the scarcity of very metal poor stars in the halo,  the new very metal poor candidates are faint and require a significant fraction of observing time on 10 m class telescopes.  
 Consequently, the number of stars  with 
 [Fe/H] $\simeq  -3.5$  for which  the Eu abundance has been measured is small. 
 The sample used in the present paper does not represent an unbiased sample of stars, so the distribution function of [Eu/Fe] vs [Fe/H] has no significance. In several stars, 
 the upper limit of the Eu abundance has been computed. These upper limits are useful as they complement the real measurement and show that, so far, no star below [Fe/H] $\simeq  -3.5$  has
 a very high [Eu/Fe] ratio, i.e. of the order of what is found in stars at [Fe/H] $\simeq  -3.5$. Although 
we cannot fully rule out the existence of such stars with [Eu/Fe] $\ge$ 1.2 dex,  the present sample seems 
to indicate that the upper envelope of the [Eu/Fe] stops to rise at [Fe/H]  $\simeq  -3.0$ 
This hypothesis receives support from the recent study of \citet{hansen14} who analyzed 4 stars with [Fe/H] 
$\leq -4.00$ and did not report any measurement of the Eu abundance although the wavelength
range  of the spectra used in this study covers the Eu transitions wavelength.

\begin{figure*}
\begin{center}
\epsfig{file=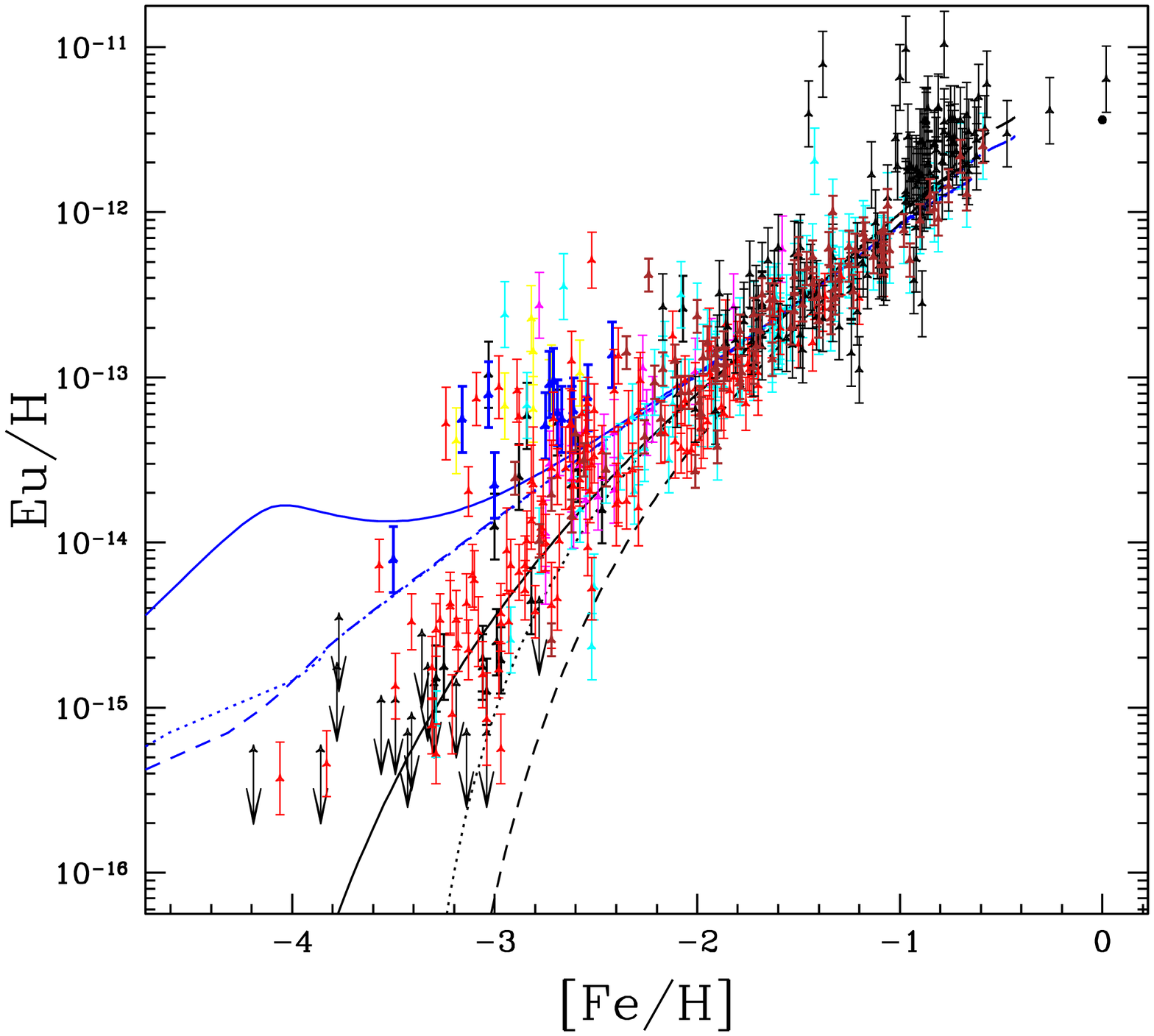, width=0.45\linewidth}
\epsfig{file=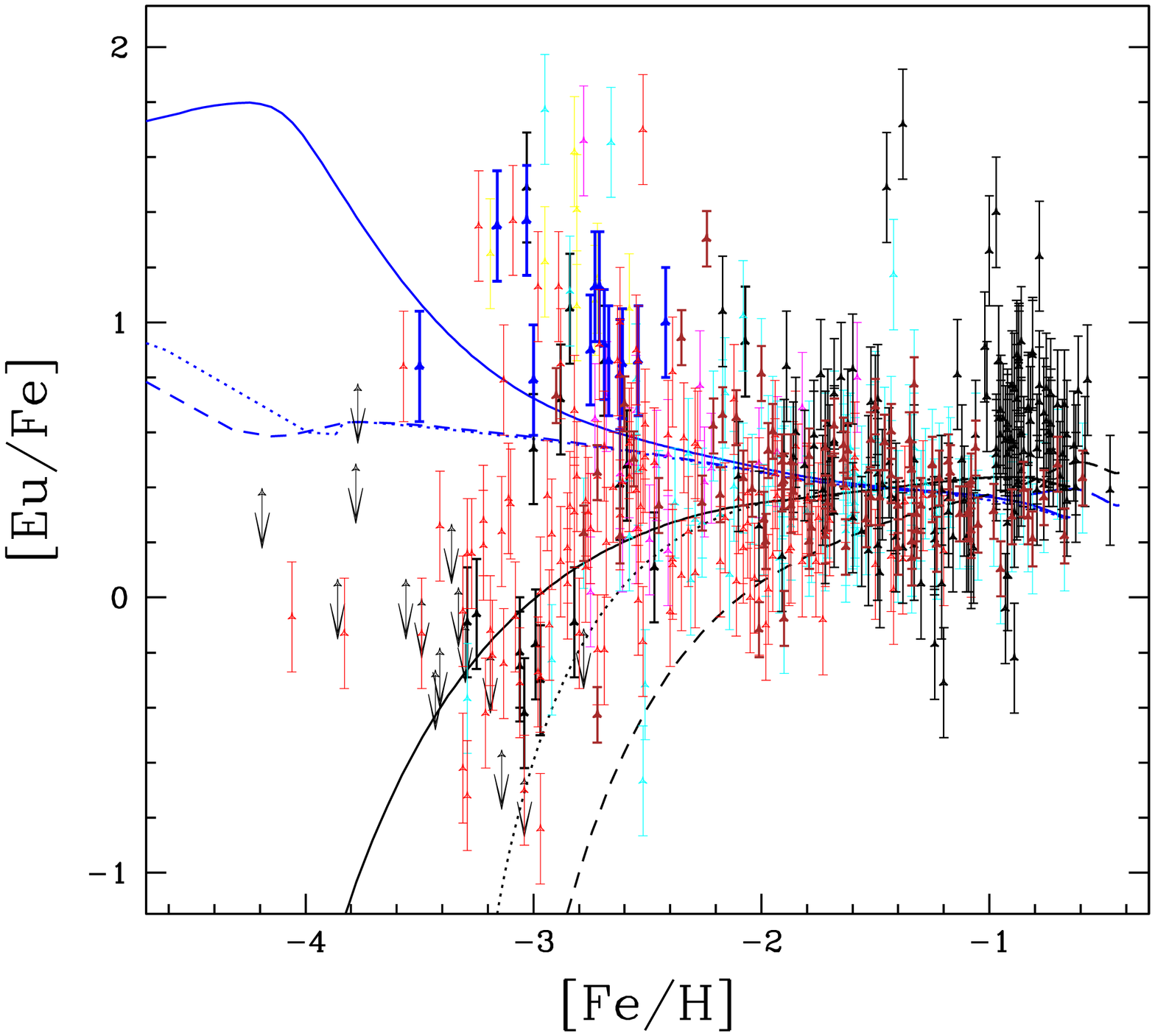, width=0.45\linewidth}
\end{center}
\caption{\textbf{Cosmic evolution of Eu: comparison between the two possible astrophysical sites Eu/H and [Eu/Fe] vs [Fe/H].}
Evolution of Eu/H  and [Eu/Fe] as a function of  [Fe/H] either in the CCSN (blue lines) or in the NSM (black lines) scenario.
The evolution is computed for the three SFR modes considered in the present study, SFR1 (solid lines), SFR2 (dotted lines), SFR3 (dashed lines). In the CCSN scenario, the Eu yield is  $10^{-7}~M_\odot$ per supernova. In the NSM scenario, the Eu yield is  $7 \times10^{-5}~M_\odot$ per merger, the fraction of binary compact objects is $\alpha= 0.002$ and the coalescence timescale is $\Delta t_\mathrm{NSM}=0.2\, \mathrm{Gyr}$. 
Data points come from different references: black points and upper limits at low metallicity from \citep{francois07}, brown points at higher metallicity from \citep{sim04}, yellow from \citep{barklem05}, magenta from \citep{ren12}, red from \citep{roed10, roed14a, roed14b}, heavy blue points from \citep{roed14c}. 
The bulk of black points at high metallicity come from dwarf spheroidal systems \citep{shertone01, shertone03, geisler05, cohen09, letarte10, Starkenburg13, McWilliam13}.
}
\label{fig:EuH}
\end{figure*}

The inspection of observed abundances (see Fig.~\ref{fig:EuH})  shows the apparent existence of two branches  starting  below [Fe/H] $\simeq -2.5$, the first one with high values of [Eu/Fe] as high as +2.0, the second one with decreasing values down to [Eu/Fe] $\simeq  -1.0$ at [Fe/H] $\simeq  -3.2$. As the sample is biased, it is difficult
to be fully confident about the reality of this two branches.  The plot could also well be interpreted as an 
increase of the dispersion as the metallicity decreases.
 Indeed, semi analytic models of merger tree which trace the chemical evolution of matter in different regions of the Universe (from over- to under-dense regions) show clearly a dispersion growing as a function of the redshift \citep{dvorkin15}.

The origin of the stars with a high level of  r-process elements like Eu is still a matter of debate. A recent analysis by \citet{roed14b, roed14c} has shown and confirmed that this overabundance is also found in main-sequence turn-off stars rejecting the hypothesis that the star would have  produced itself such overabundances.  They also did not find any compelling evidence to suggest that a noticeable high fraction of highly r-process enriched stars are members of binary systems, assigning the origin of  the r-process  enhancement to a companion star. 
 They also confirmed that these peculiar stars do not present a different $\alpha$-element chemical signature from the bulk of the other metal poor stars. Therefore, the site responsible for the production of this r-process enhancement is not expected to produce any chemical anomalies for light elements.

\begin{figure*}
\begin{center}
\epsfig{file=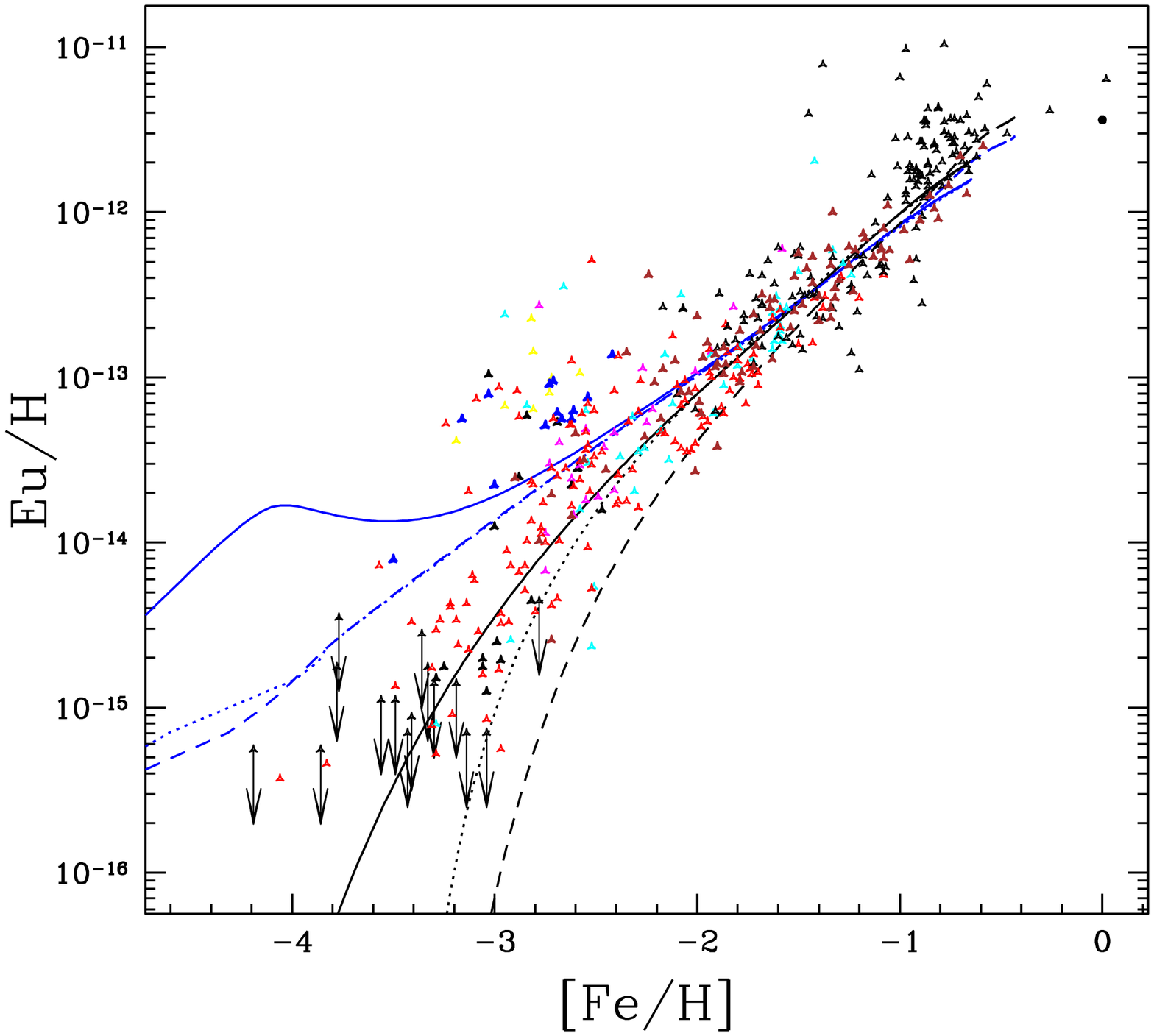, width=0.45\linewidth}
\epsfig{file=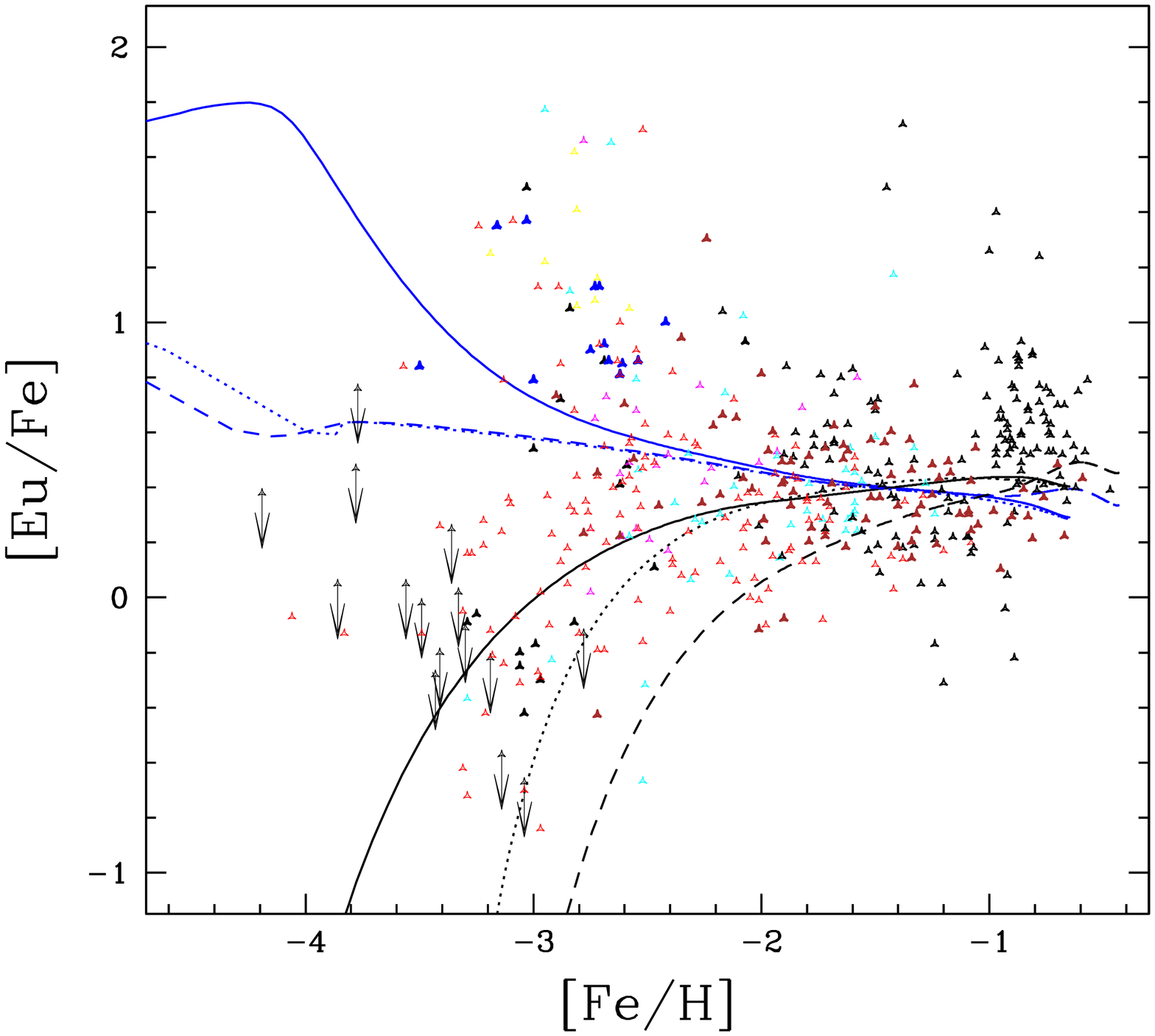, width=0.45\linewidth}
\end{center}
\caption{ Same as Fig.~\ref{fig:EuH} but without error bars.}
\label{fig:EuHb}
\end{figure*}

\subsection{Eu: yields from CCSN and NSM}
\label{sect_site}

Each of the two possible r-process sites, CCSN and NSM, may potentially contribute to the galactic r-process enrichment, but we will assume in the present work that they are mutually exclusive, so that the the r-process in the Universe is either originating from CCSN or NSM but not from a mix of both sites. While the cosmic evolution of r-process elements in the CCSN scenario essentially depends on the amount of r-process material ejected by the explosion, in the NSM scenario, it is also function of the binary fraction of compact objects $\alpha$ and the coalescence timescale of the binary system $\Delta t_\mathrm{NSM}$. We describe here the adopted yields of Eu expected to be ejected from CCSN or NSM.

\subsubsection{Eu yields from CCSN}

Since there exists so far no consistent r-process calculation in CCSN, the Eu yields from supernovae remain highly uncertain. The CCSN Eu yields are therefore free parameters that can only be adjusted to reproduce some observables, like the [Eu/Fe] trend as a function of the [Fe/H] observed in galactic stars or the total amount of Eu material in the Galaxy. In particular in this latter case, if we assume that the heavy-element composition in the solar system is representative for the whole Galaxy (as suggested by the small scatter of the Eu abundance in the present-day Milky Way), for a Eu mass fraction of $3.7\times 10^{-10}$ \citep{lod03,asplund09} and a total galactic baryonic mass of $6 \times 10^{10}~M_{\odot}$ \citep{millan11}, this implies a Eu Galactic content of $22~M_{\odot}$. Assuming that the CCSN events occur at a constant rate identical to the currently observed one, i.e. about one per century during the approximate age of the Galaxy of $1.3~ \times 10^{10}$~yr, some $1.3~10^8$ type-II explosions should be responsible for the $22~M_{\odot}$ of Eu, i.e. each CCSN should in this case produce about $1.7 \times 10^{-7}~M_{\odot}$ of Eu. For this reason, by simplicity, a constant yield of $10^{-7}~M_{\odot}$ of Eu is adopted here for all CCSN, regardless of the mass of the progenitor.

\subsubsection{Eu yields from NSM}
\label{sect_euy}

While a successful r-process cannot be obtained {\it ab initio} in CCSN simulations, NSM models consistently predict a significant production and ejection of r-process material. In particular, a mass-weighted combination of the dynamical ejecta from both the binary merger phase and the secular ejecta from the BH-torus evolution has been shown \citep{just14} to reproduce the solar r-abundance pattern, and therefore also the elemental one seen in ultra-metal-poor stars, amazingly well in the whole $90 \la A \le 210$ range. This calculation represents today the most consistent estimate of Eu production by NS-NS and NS-BH mergers taking consistently into account both the prompt ejecta as well as the relic BH-torus outflow contributions. For this reason, the Eu yields obtained in such combined models are considered in the present study.  More precisely, the three combined systems (merger model plus remnant model) corresponding to models with torus masses $M_\mathrm{torus} = 0.03, 0.1$ and $0.3~M_{\odot}$ \citep[see Fig. 19 of][]{just14} are found to eject between $7 \times 10^{-5}$ and $2 \times 10^{-4}~M_{\odot}$ of Eu. A fiducial Eu yield of $7 \times 10^{-5}~M_{\odot}$ is adopted here (see the distribution of the ejected mass in NSM mergers in Fig. 6 in \cite{Fryer15}).

\subsection {Eu: cosmic evolution}
\label{sect_res0}

\subsubsection{Evolution of Eu as a function of [Fe/H]}
\label{sec:standard}
The Eu cosmic evolution is 
computed for the three SFR modes.
Figs.~\ref{fig:EuH} and \ref{fig:EuHb} (with and without error bars)
 show the predicted evolution of Eu/H and [Eu/Fe] as a function of [Fe/H] for both r-process sites, separately, the CCSN (blue lines) and NSM (black lines), with a comparison to the observations described in Sect.~\ref{sect_obs}. 

Concerning the CCSN scenario, the Eu yield of $10^{-7}~M_\odot$ is seen to reproduce fairly well the spectroscopic data for ${\rm [Fe/H]} > - 2.5$. At lower metallicity, Eu is clearly overproduced by CCSN, specifically if we consider
Eu data at lowest metallicity, 
as pointed in  Sect.~\ref{sect_obs}. In the SFR1 mode, where a Pop III contribution is included, the overproduction is even more significant due to the lack of iron production at very low metallicities in the models of \citet{ww95} (Case B),  whereas Eu is assumed to be produced in all mass ranges of massive stars. 

\begin{figure*}
\begin{center}
\epsfig{file=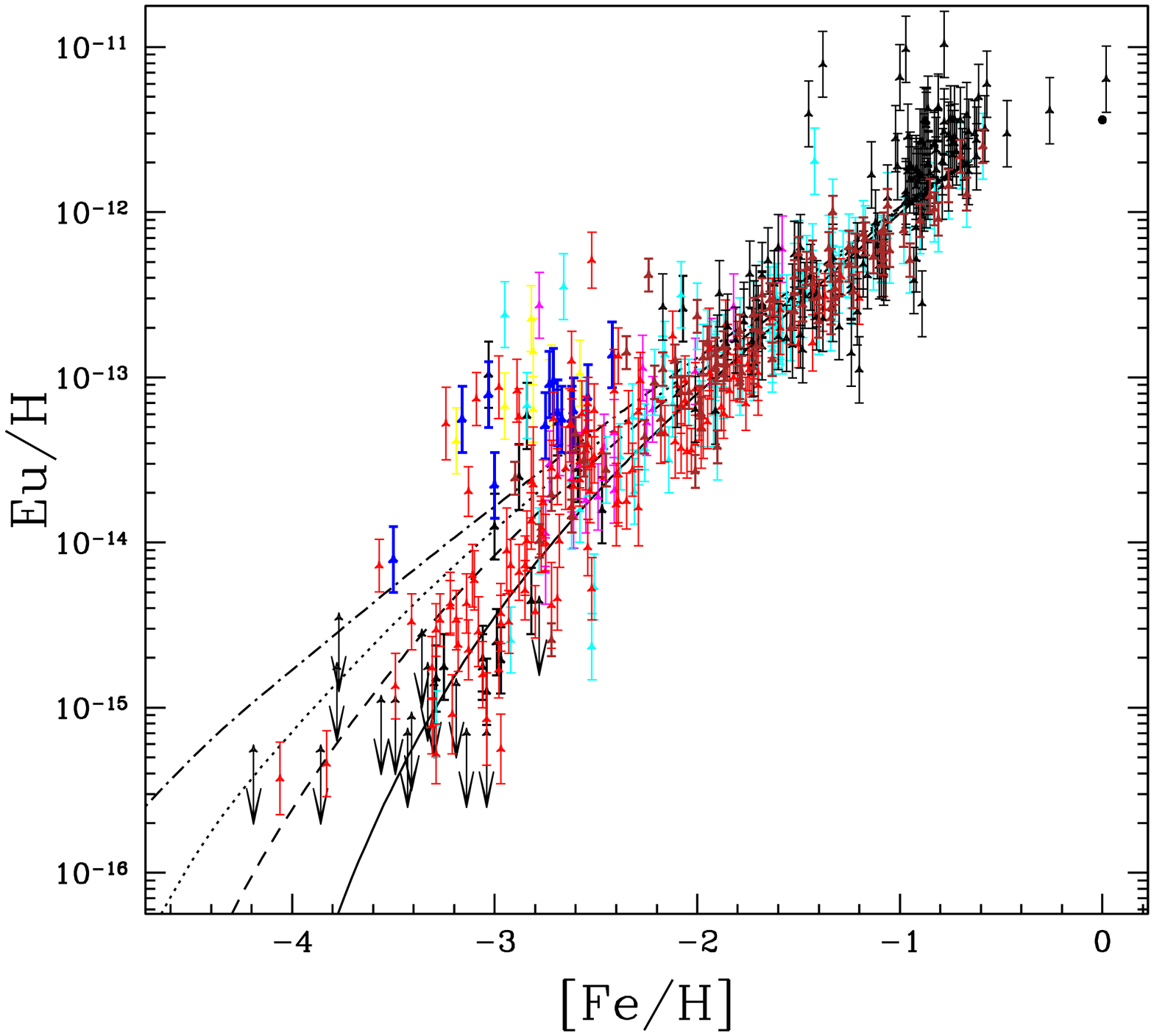, width=0.35\linewidth}
\epsfig{file=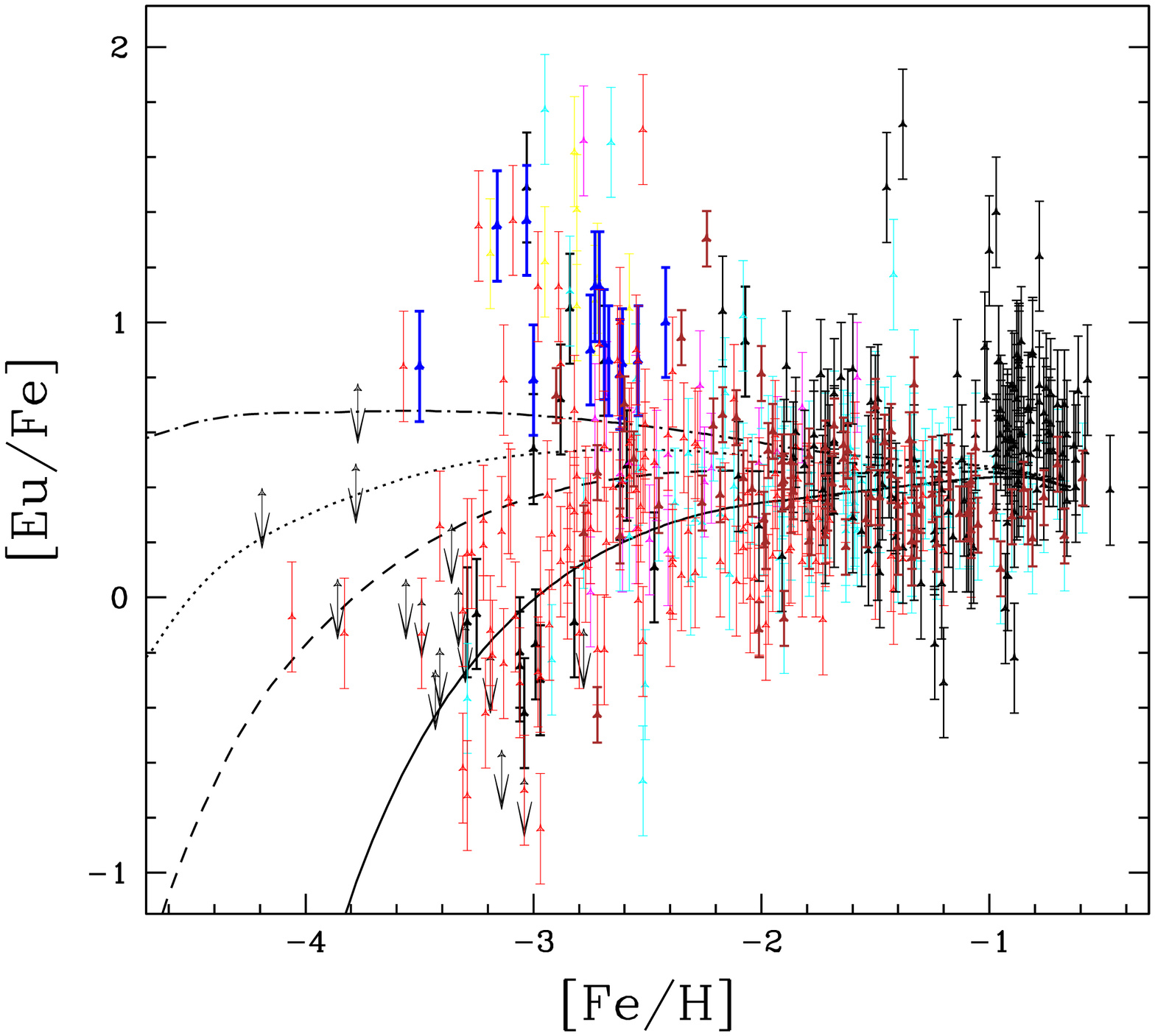, width=0.35\linewidth}
\end{center}
\caption{\textbf{Cosmic evolution of Eu in the NSM scenario: effect of the coalescence timescale (SFR1 case).}
Same as Fig.~\ref{fig:EuH} in the NSM scenario and SFR1 case. 
We compare four different coalescence timescales: $0$ (dot-dash line), $0.05$ (dotted line), $0.1$ (dashed line) and $0.2\, \mathrm{Gyr}$ (solid line).
}
\label{fig:EuFeSFR1}
\end{figure*}

\begin{figure*}
\begin{center}
\epsfig{file=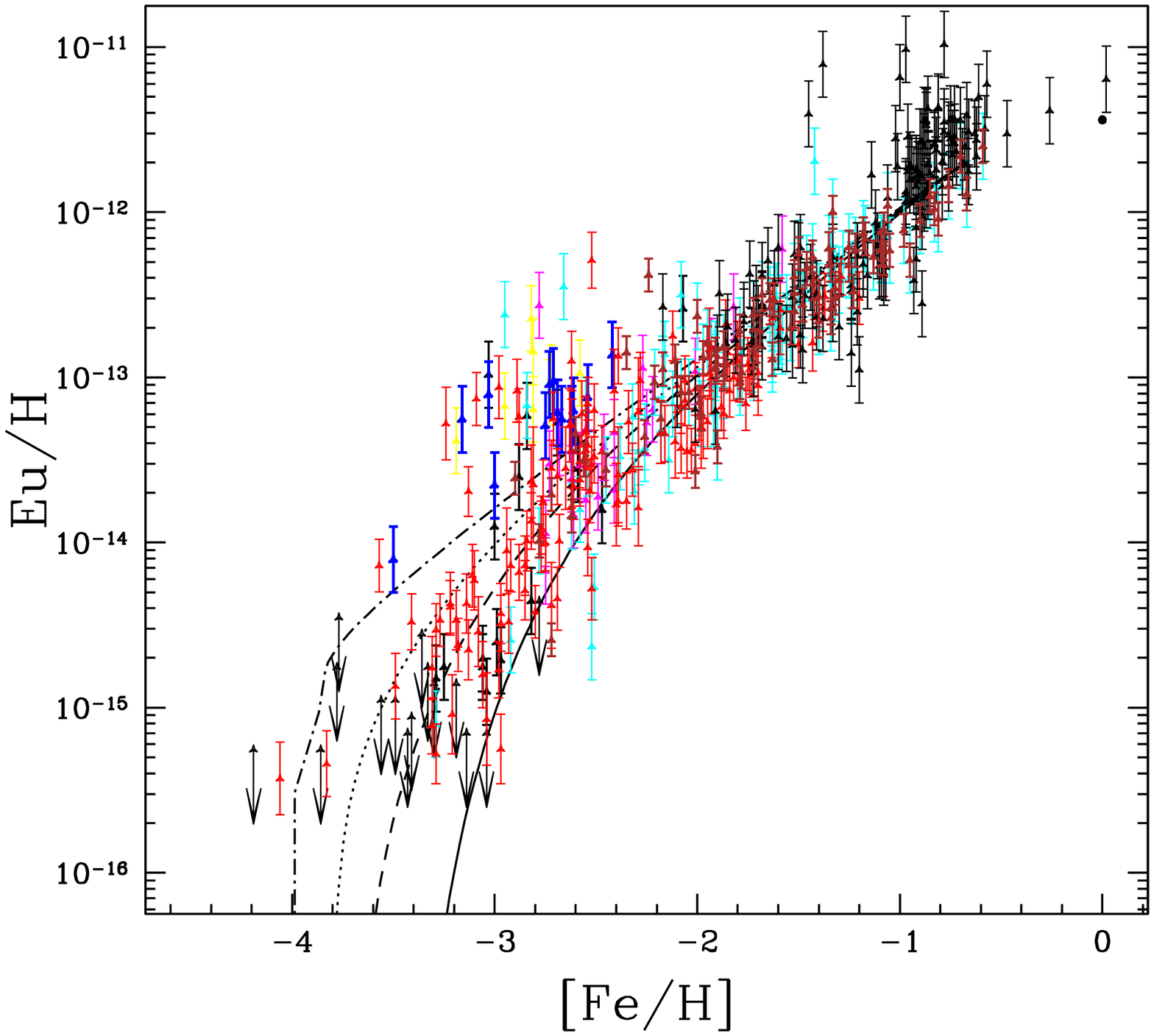, width=0.35\linewidth}
\epsfig{file=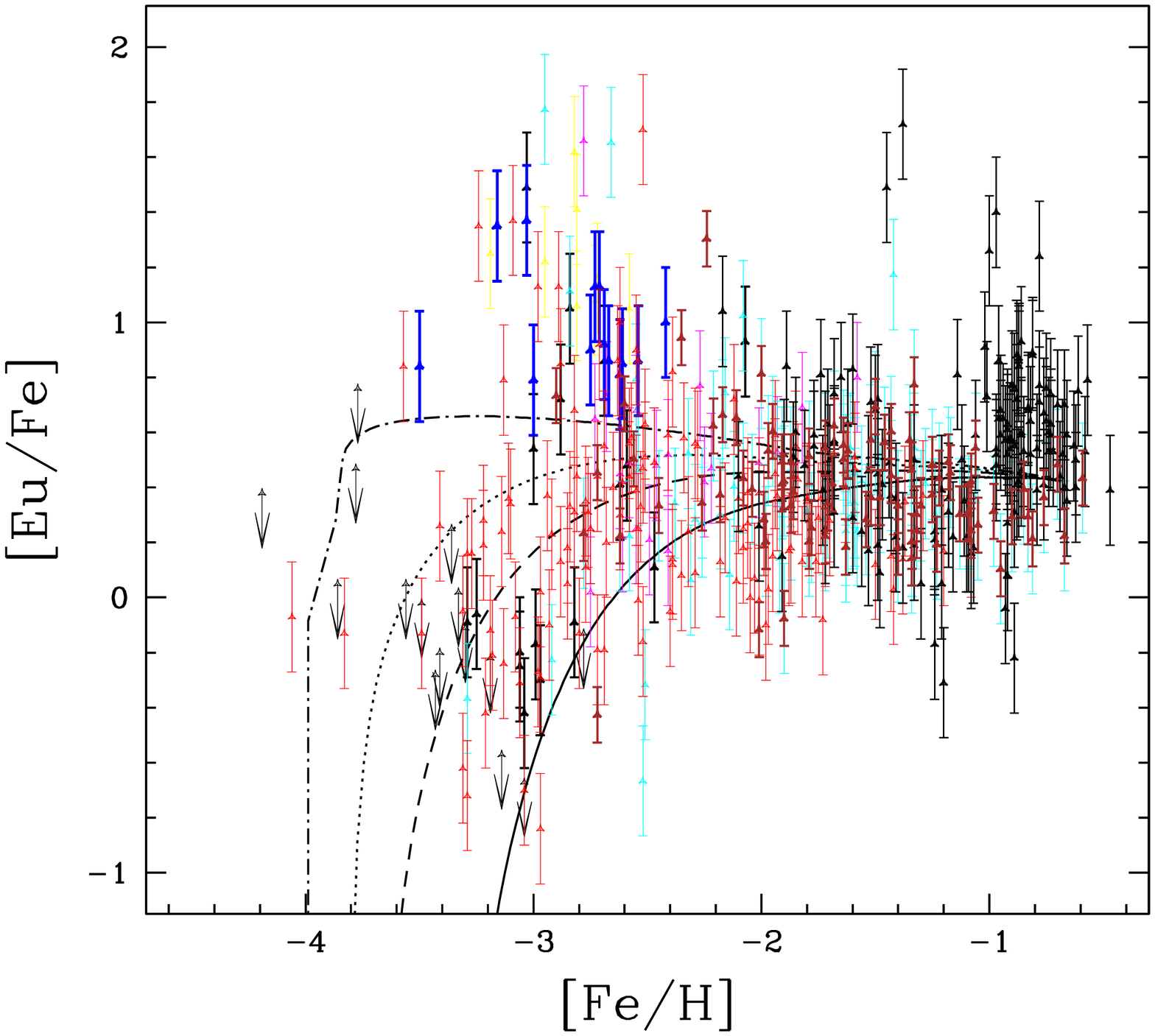, width=0.35\linewidth}
\end{center}
\caption{\textbf{Cosmic evolution of Eu in the NSM scenario: effect of the coalescence timescale (SFR2 case)}.
Same as Fig.~\ref{fig:EuFeSFR1} in SFR2 case. 
}
\label{fig:EuFeSFR2}
\end{figure*}

\begin{figure*}
\begin{center}
\epsfig{file=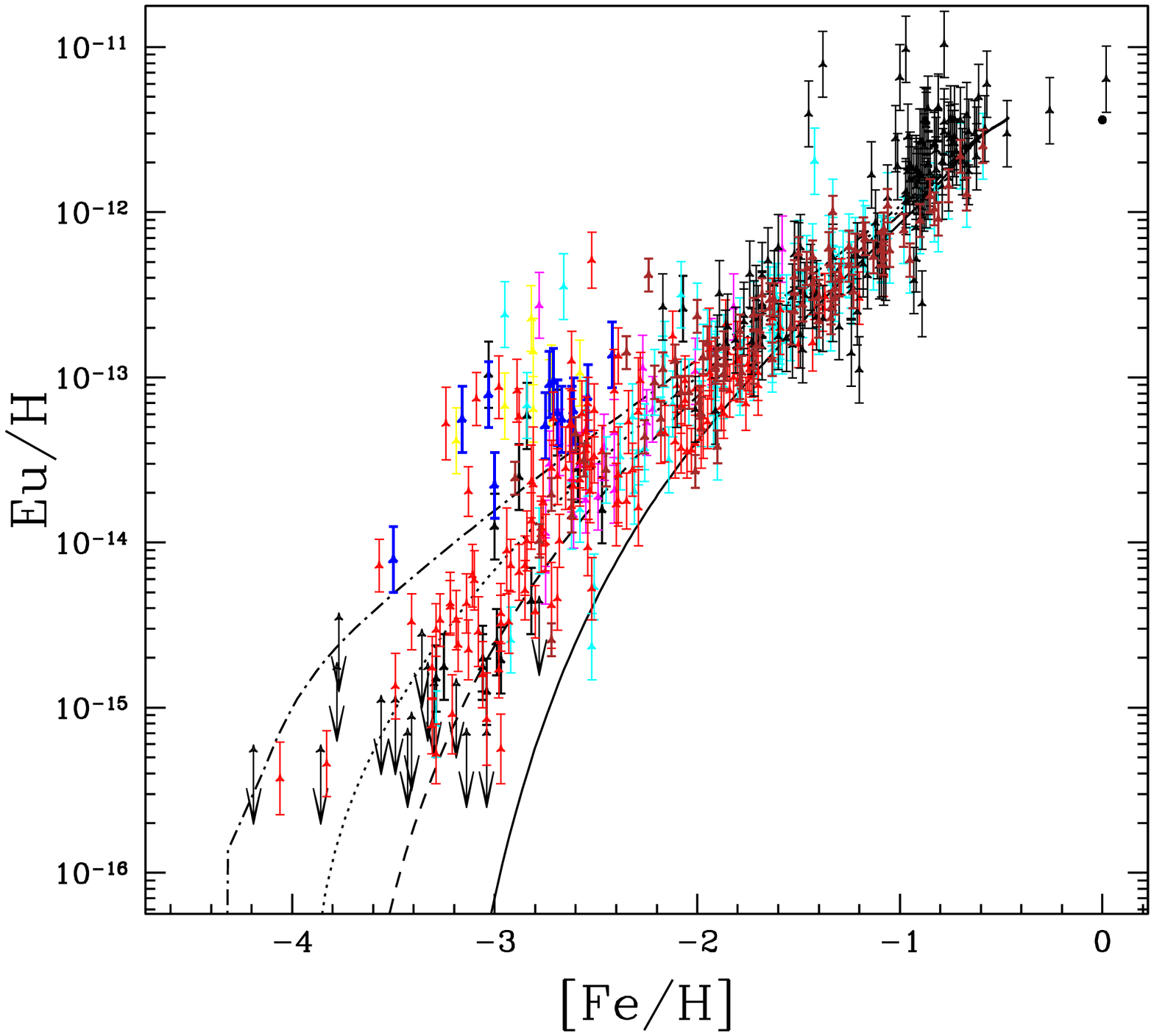, width=0.35\linewidth}
\epsfig{file=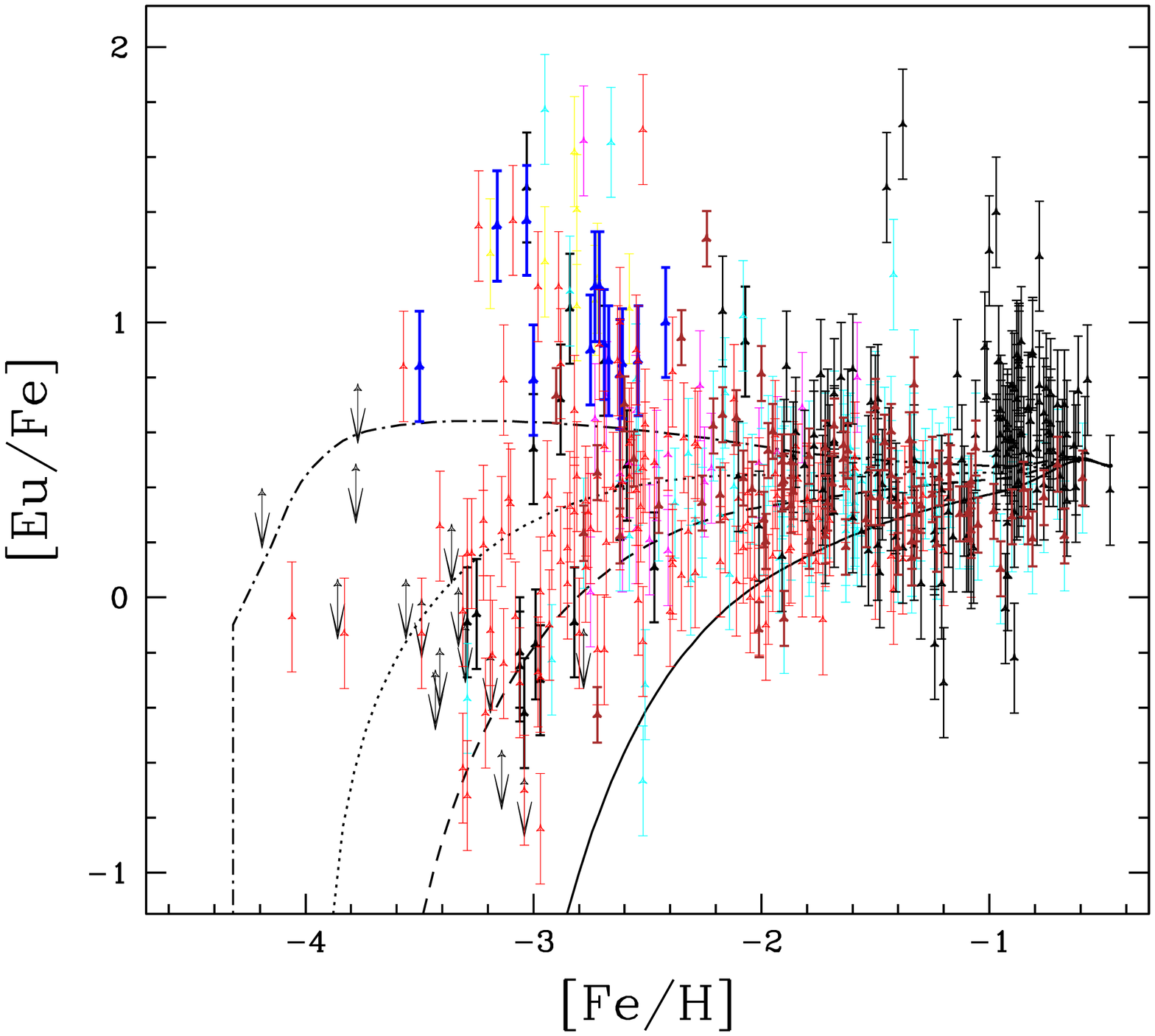, width=0.35\linewidth}
\end{center}
\caption{\textbf{Cosmic evolution of Eu in the NSM scenario: effect of the coalescence timescale (SFR3 case).}
Same as Fig.~\ref{fig:EuFeSFR1} in SFR3 case. 
}
\label{fig:EuFeSFR3}
\end{figure*}

Regarding the NSM scenario, with the fixed value of $7 \times 10^{-5}~M_\odot$ for the Eu yield (see \S~\ref{sect_euy}), observations at ${\rm [Fe/H]} > - 2.5$  are seen to be in good agreement with 
the predicted evolution
 for a fraction of binary compact objects $\alpha= 0.002$
and a coalescence timescale 
$\Delta t_\mathrm{NSM}= 0.2$~Gyr.
A larger Eu yield of $2 \times  10^{-4}~M_\odot$ would logically imply a lower value of $\alpha \simeq 0.0007$, both parameters being degenerate in this approach. 
The
 agreement with the observations of Eu at low metallicity depends strongly on the coalescence timescale, which can therefore be constrained for each SFR mode, as illustrated in Figs.~\ref{fig:EuFeSFR1}, \ref{fig:EuFeSFR2} and~\ref{fig:EuFeSFR3}, where we 
 compare four coalescence time scales
 $\Delta t_\mathrm{NSM}=0$, 0.05, 0.1, 0.2~Gyr for each SFR mode. 
The best fit is obtained for a typical coalescence timescale of 0.2 Gyr for SFR1 (Fig.~\ref{fig:EuFeSFR1}), 0.1 Gyr for SFR2 (Fig.~\ref{fig:EuFeSFR2}) and 0.05 Gyr for SFR3 (Fig.~\ref{fig:EuFeSFR3}).

The difference
stems from the [Fe/H] evolution which differs significantly for the three SFR modes, as already pointed out in \S~\ref{sect_in} (see Fig.~\ref{fig:iron}). The more iron produced at early time, the shorter coalescence timescale needed to explain the low-metallicity r-process-rich stars. 
An important conclusion, illustrated by  these results, is that a coalescence timescale larger than typically 0.3~Gyr is excluded by the observations of Eu at low [Fe/H] for all SFR.
  Note that the production of iron from type Ia supernovae is included in our calculation but their contribution becomes efficient only at the end of the cosmic evolution,  
i.e. for ${\rm [Fe/H]} \ga -1$. Indeed, due to the lifetime of the $2\,M_\odot$ stars (2 Gyr) and the potential time delay of explosions, SNIa iron enriches the ISM at $z <3$.
It should also be noted that, contrary to the CCSN case, the Pop III component in the SFR1 mode does not affect the predictions in the NSM scenario due to the time delay of the merging.


\subsubsection{Evolution of Eu as a function of redshift}

\begin{figure}
\begin{center}
\epsfig{file=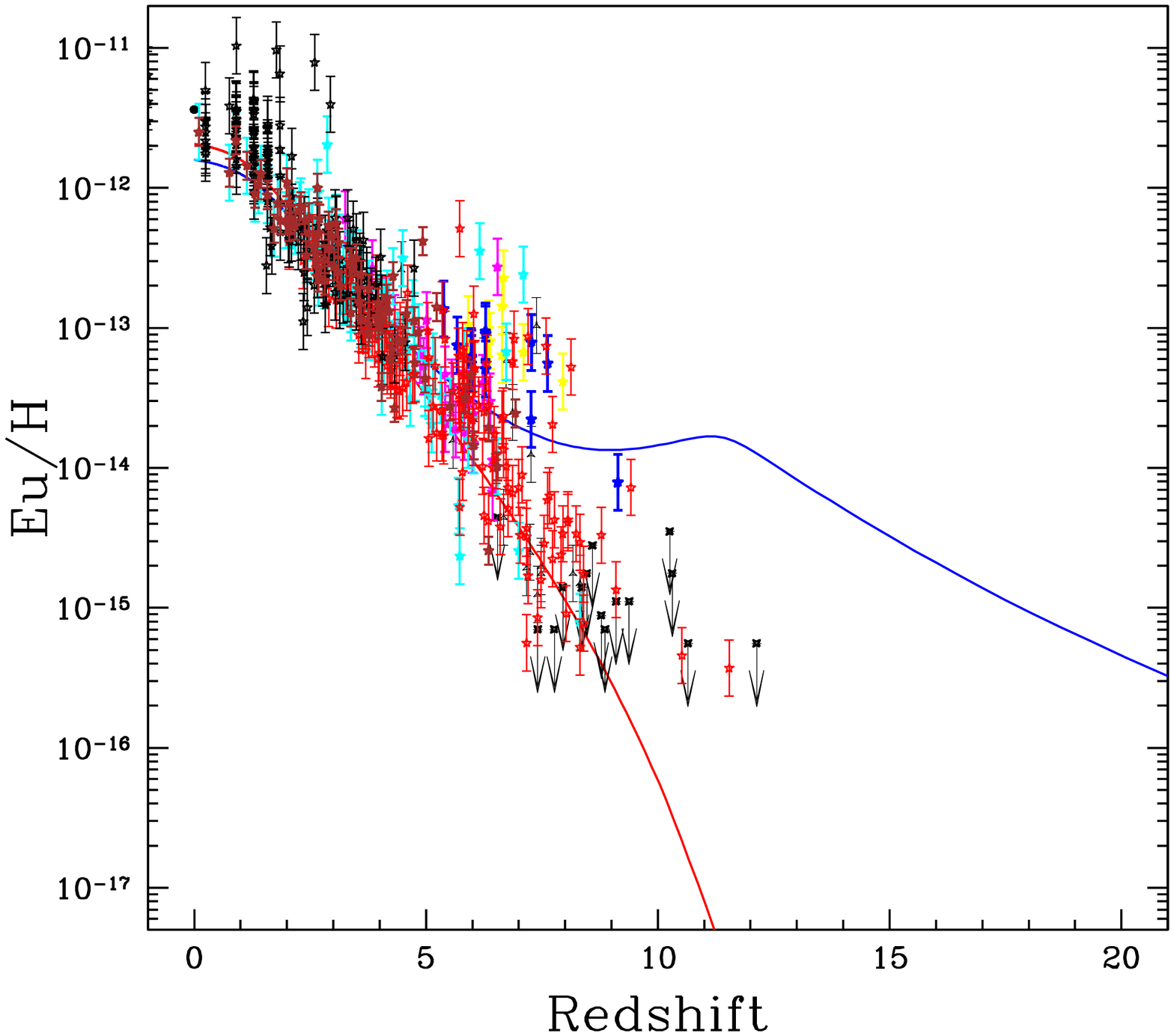, height=2in}

\epsfig{file=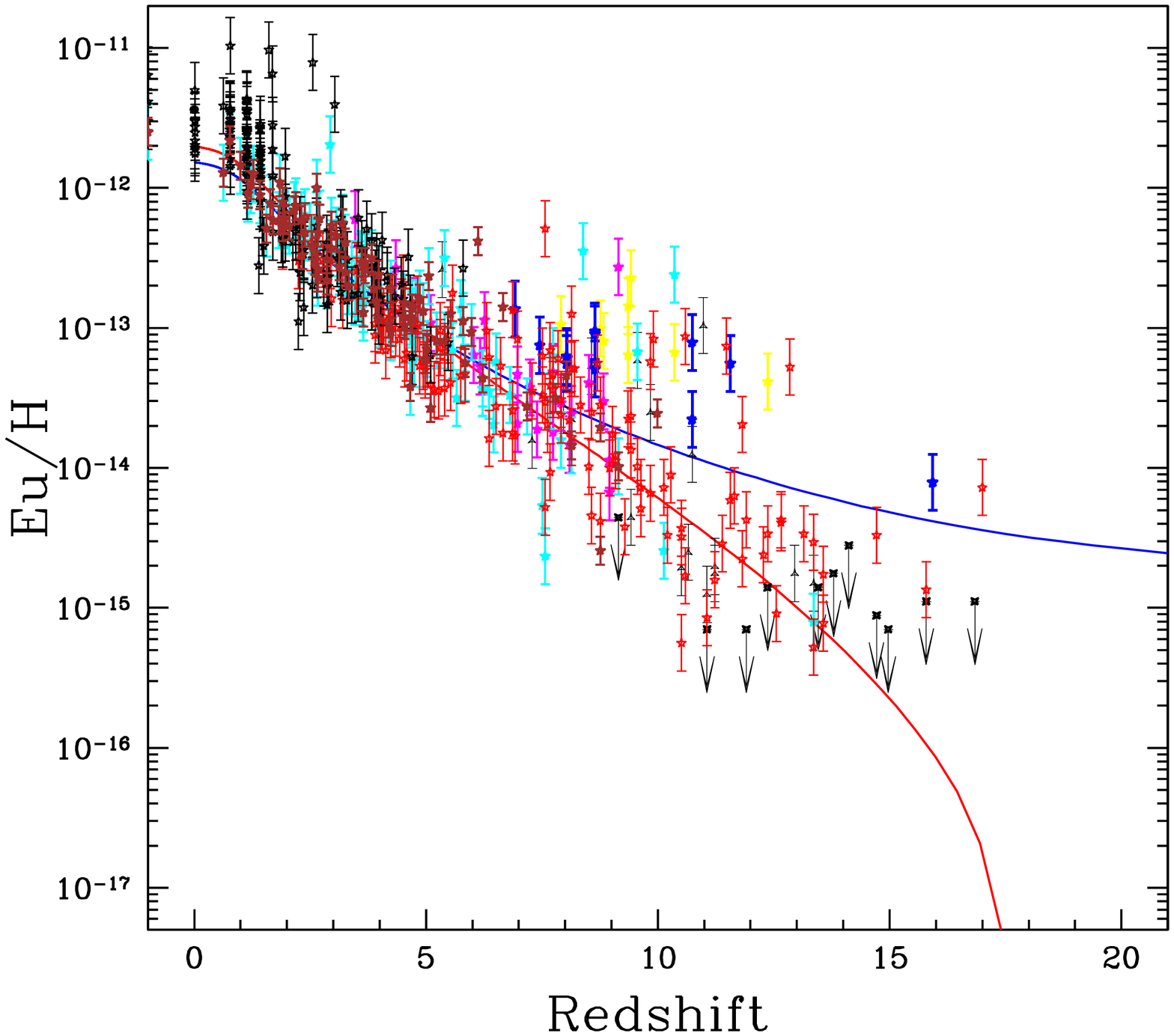, height=2in}

\epsfig{file=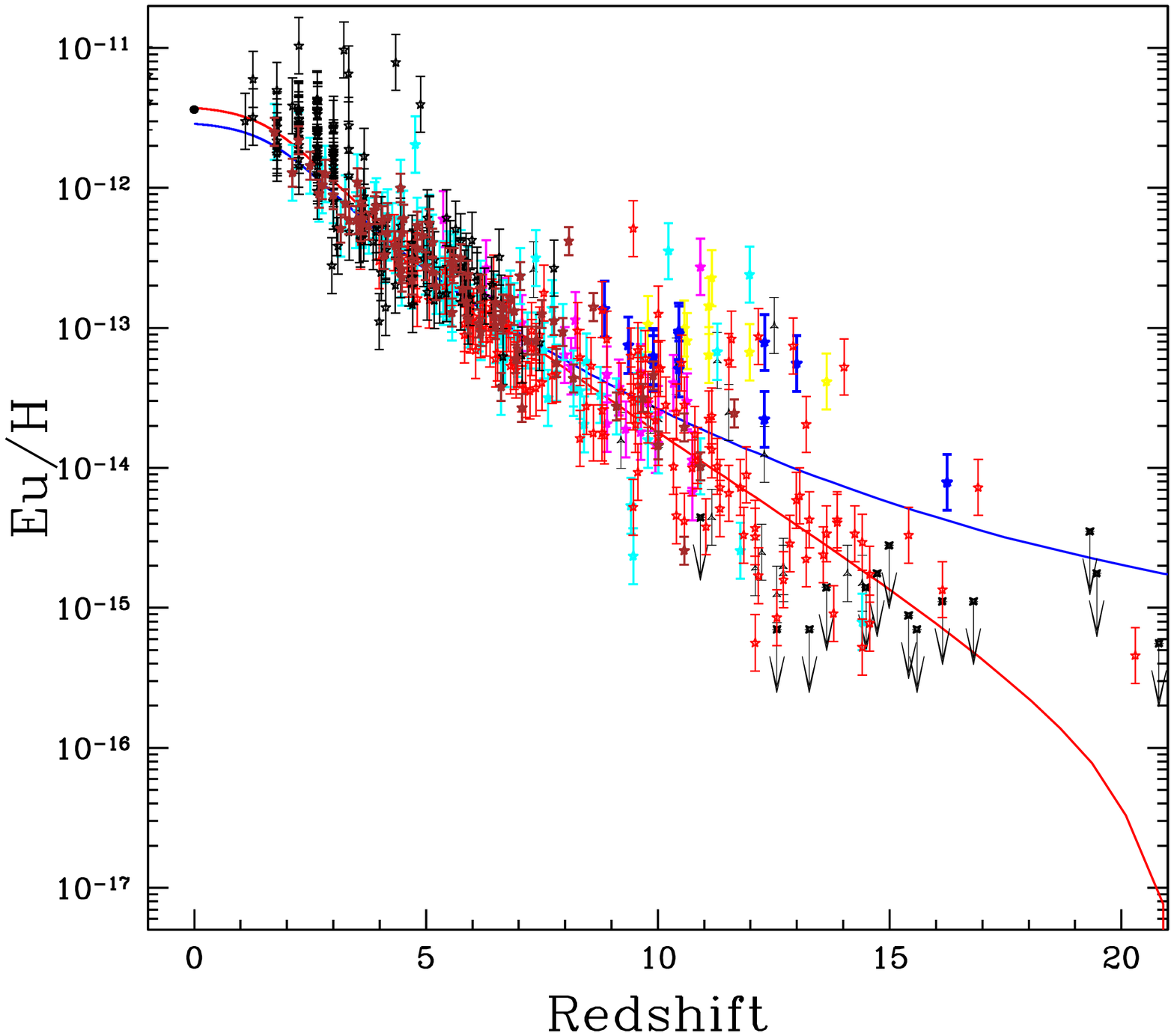, height=2in}

\end{center}
\caption{
\textbf{Cosmic evolution of Eu: comparison between the two possible astrophysical sites (2) Eu/H vs redshift.}
Evolution of Eu/H  as a function of  $z$ either in the CCSN (blue lines) or in the NSM (red lines) scenario.
The evolution is computed for the three SFR modes considered in the present study, SFR1 (upper panel), SFR2 (middle), SFR3 (lower panel). 
The three SFR modes are shown in different panels because the observations expressed in terms of [Fe/H] do not correspond to the same redshift $z$ when different SFR are considered.
The yields in the two scenarios are the same as in Fig.~\ref{fig:EuH}. In the NSM scenario, the binary fraction is $\alpha=0.002$ and the coalescence timescale is $0.2$~Gyr for the SFR1 mode, and $0.1$~Gyr for the SFR2 and $0.05$~Gyr for SFR3.}

\label{fig:EuHbis}

\end{figure}

Fig.~\ref{fig:EuHbis} shows the Eu evolution
 as a function of redshift for the three SFR modes. To include observational data as a function of $z$,  each [Fe/H] value needs to be associated with a corresponding redshift. In our approach, the iron abundance increases monotonically with time, so that there exists a unique relation between [Fe/H] and $z$, using equation 1, but this relation is however function of the adopted SFR since the Fe evolution depends on the SFR, as shown in Fig.~\ref{fig:iron} (lower panel). For this reason, the three modes are shown in different panels.
In the NSM scenario, the coalescence timescale adopted for each SFR mode is $\Delta t_{NSM}= 0.2$, 0.1, 0.05 ~Gyr  for SFR1, SFR2, SFR3, respectively, as discussed above.

\subsection{Sensitivity analysis}

\subsubsection{Eu from CCSN and the Pop III star component}

\begin{figure}
\begin{center}
\epsfig{file=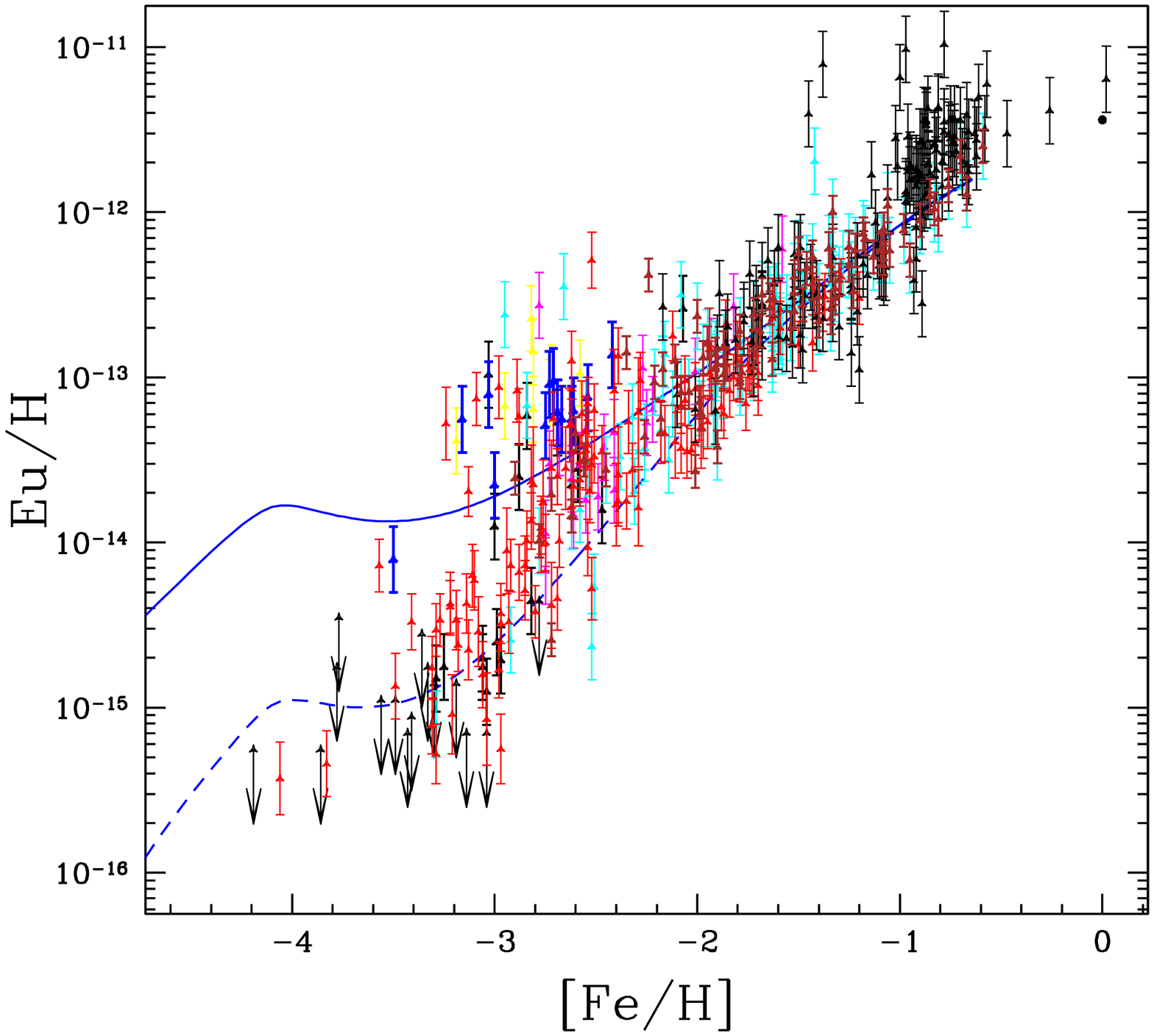, width=0.9\linewidth} 
\end{center}
\caption{\textbf{Cosmic evolution of Eu in the CCSN scenario: effect of the Eu yield (SFR1 case).}
Evolution of Eu/H  as a function of  [Fe/H]  in the CCSN  scenario for the SFR1 mode, assuming either the standard prescription for the Eu yield (solid line), i.e. $10^{-7}~M_\odot$ per supernova for massive stars at all metallicities, or a modified prescription (dashed line) where Eu is produced in supernovae only for massive stars with a metallicity  $Z >10^{-4} Z_\odot$.}
\label{fig:EuHter}

\end{figure}

As already mentioned, the Eu nucleosynthesis by CCSN remains highly uncertain. In our standard calculations (\S~\ref{sect_res0}), it was assumed that each CCSN produces $10^{-7}~M_\odot$ of Eu regardless of the mass and metallicity of its progenitor. If we consider in the SFR1 mode the existence of a Pop III star component, as required to explain the 
Thomson optical depth of the CMB measured by
WMAP9 (see Fig.~\ref{fig:ionisation}), Fig.~\ref{fig:EuH} clearly shows a large overproduction of Eu at low metallicity. So, would the r-process successfully take place in CCSN, the observation of Eu in ultra-metal-poor star could be a discriminant indicator highlighting the existence or not of Pop III stars. However, if we assume that Pop III stars, or more generally $Z < 10^{-4} Z_\odot$ stars, do not produce r-process elements significantly (less than typically 
$5~10^{-9}~M_\odot$), the Eu evolution curve in SFR1 model, becomes almost equivalent to the predictions obtained with SFR2 and SFR3, as shown in Fig.~\ref{fig:EuHter}.

\subsubsection{Sensitivity to the early iron production}

\begin{figure}
\begin{center}
\epsfig{file=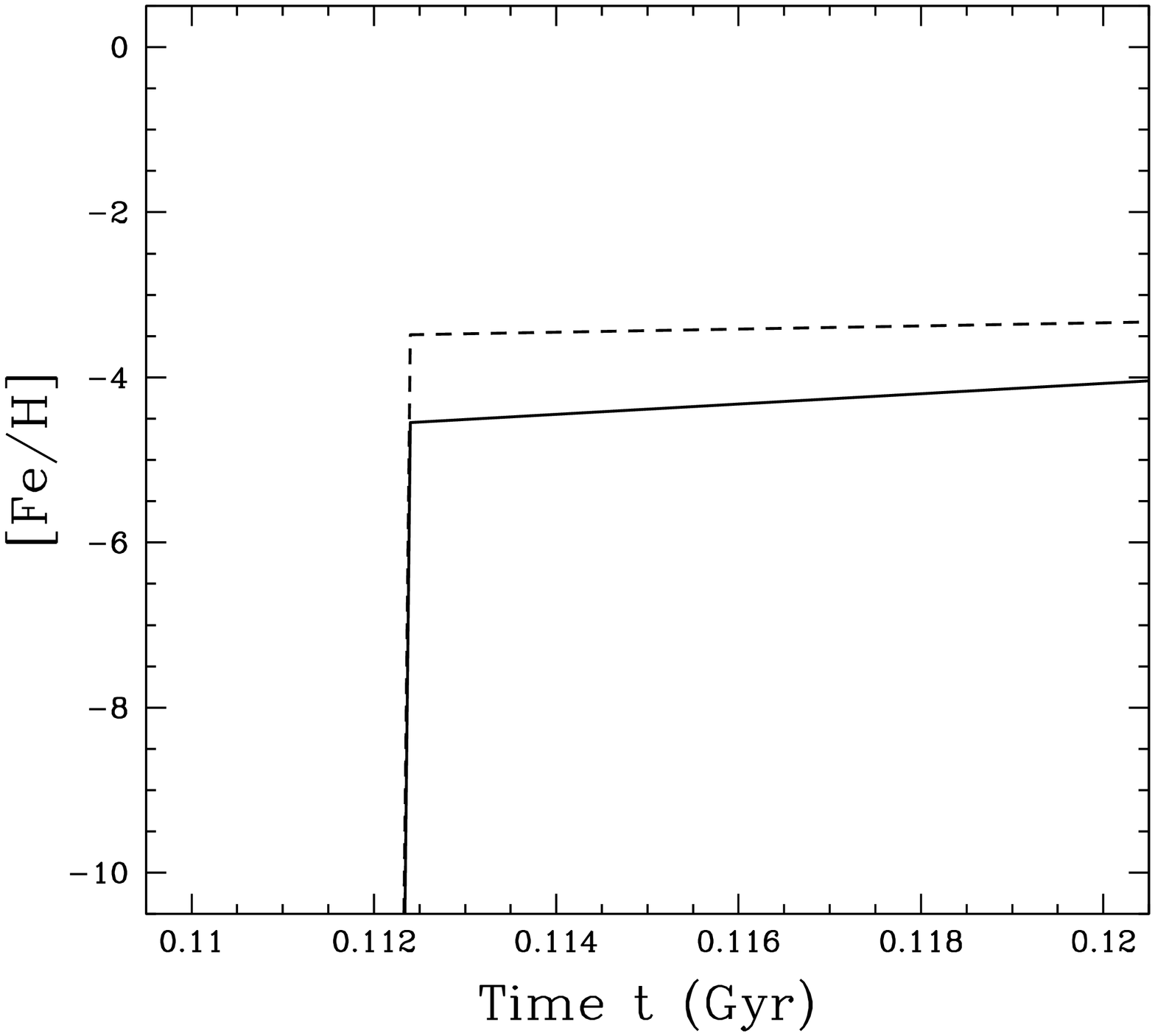, height=3in}
\epsfig{file=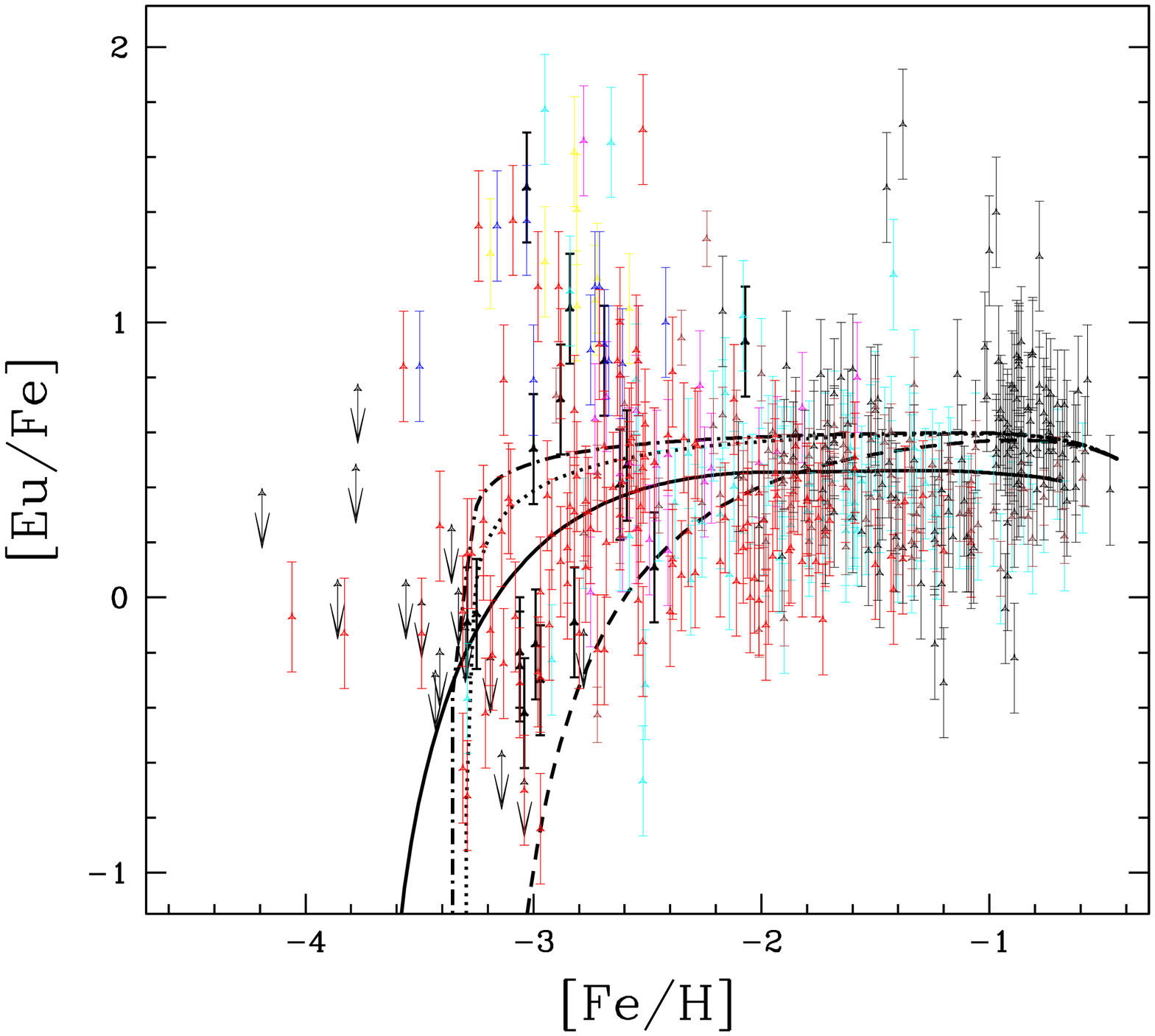, height=3in}
\end{center}
\caption{
\textbf{Cosmic evolution of Eu in the NSM scenario: sensitivity to the iron production and to the lower end of the IMF.}
Upper panel: evolution of [Fe/H] as a function of time in the SFR2 case, computed either with our standard prescription (solid line: Fe yields from \citet{ww95}, case B,  and $M_\mathrm{inf} = 0.1~M_\odot$ for the IMF), or with a modified prescription (dashed line:  Fe yields from \citet{Koba06}, hypernova case,  and $M_\mathrm{inf} = 0.8~M_\odot$ for the IMF) ;
Lower panel: corresponding evolution of  [Eu/Fe] as a function of [Fe/H]: our standard prescription for iron production (solid line) with a coalescence timescale of $\Delta t_\mathrm{NSM}=0.1$~Gyr is compared with the modified prescription for three  values of the time delays, i.e. $0.1$~Gyr (dashed line),  $0.01$~Gyr (dotted line), $0.001$~Gyr, (dot-dash line).
}

\label{fig:test1}
\end{figure}

Our results depend on the adopted scenario for star formation in the Universe. This is why a detailed sensitivity study has been performed with different SFR stories, different stellar yields, with the aim of fitting the maximum of observational constraints, i.e. cosmological and local.

Recent studies published by other authors \citep{mat14, kom14}
concluded that a merger time scale of the order of 1 - 10 Myr was necessary to reproduce 
the observed evolution of r elements with NS mergers alone. 
We find a different conclusion constraining the merger time scale to the range of ~100 - 200 Myr. 
We made a careful analysis of the origin of this discrepancy. We show that the merger timescale 
depends strongly on the iron evolution which itself depends on the global astration rate and the choice of yields in supernovae or in hypernovae. 
Unfortunately, the CCSN production of Fe is poorly known. 
It depends essentially on the explosion energy and on the mass cut (fraction of the mass trapped in the remnant).
Iron ejection is more important for higher energies of the explosion.
Typically, we choose the SNII models by \citet{ww95}  ($E_{explod.} = 10^{51}$ ergs). At solar metallicity and  for stars of 40 (20) $M_\odot$, they give an iron yield  of 0.03 (0.02) $M_\odot$ respectively and 0.02 (0.1) $M_\odot$ respectively at Z = 0.001. 
On the other hand, hypernova models by \citet{Koba06} ($E_{explod.} = 10 - 30 \times 10^{51}$ ergs)
 correspond to 0.277 (0.035) $M_\odot$ of iron for the same stars at solar metallicity and 0.26 (0.08) $M_\odot$respectively at Z = 0.001. Indeed,
hypernovae are much more energetical than the SNae. As seen in Fig.~\ref{fig:test1}, a higher Fe early production 
constraints the  merger time scale to be shorter. 
The studies by \citet{mat14, kom14}  assume stellar yields from \citet{Koba06} (hypernova case), 
whereas we assume the stellar yields from \citet{ww95} (supernova case). We favor this assumption,  since observations of the most metal-poor stars show a low iron abundance \citep{frebel15}.
It is worth noting that this result leading to merger timescales of the order of 100 - 200 Myr 
is in better agreement with our knowledge of  the properties of the binary NS population (see Sect.~\ref{sec:mergerdelay}).

As mentioned, the cosmic Fe enrichment plays a fundamental role in the interpretation of the chemical abundance evolution. To analyse the impact of this input on the determination of the NSM coalescence timescale, the Fe yield produced by massive stars has been varied in the SFR2 case.
The standard yields of  \citet{ww95}, more specifically their case B (for which low-metallicity high-mass stars do not produce iron), have been used. Note that \citep{hw10} which follow the evolution and the explosion of zero-metallicity stars in the  $10 - 100~M_\odot$ range, obtain the same results regarding the production of iron. However, as said above, other studies find different conclusions; in particular, \citet{Koba06}  predict higher iron yields than \citet{ww95} for massive stars regardless of the metallicity. Fig.~\ref{fig:test1} illustrates the impact of  Fe yields on the Eu evolution when considering the yields of \citet {Koba06}. Note that we have also modified the lower end ($M_\mathrm{inf}$) 
of the IMF from 0.1 to 0.8~$M_\odot$ to increase the gas processing in stars and favor again the iron production.
  
With the yields of \citet{Koba06}, iron at the very beginning of the evolution is seen to be synthesized  earlier than in our standard model,  leading to an overall abundance more than 10 times higher (Fig.~\ref{fig:test1}, upper panel).  We compare in Fig.~\ref{fig:test1} (lower panel) our standard SFR2 model with $\Delta t_\mathrm{NSM}= 0.1$~Gyr (solid black line) with the results obtained with the yields of \citet{Koba06} and three different coalescence timescales, namely 0.1~Gyr (dashed line), 0.01~Gyr (dotted line) and 0.001~Gyr (dot-dash line).  Consequently, a significant reduction of the coalescence timescale ($\Delta t_\mathrm{NSM} = 1$~Myr) needs to be applied to compensate for the early increase in the Fe production in this context.
It should be recalled that it is close to the  scenario obtained by \citet{mat14}  using the same Fe yields from \citet{Koba06}  and an IMF with $M_\mathrm{inf}=0.8~M_\odot$. Clearly, the more iron is produced at early time, the shorter the predicted coalescence timescale required to explain the Eu abundance in the ultra-metal-poor stars.

\subsubsection{ Time delay distribution and dispersion in data}
\label{sect_distrib}

\begin{figure}
\begin{center}
\epsfig{file=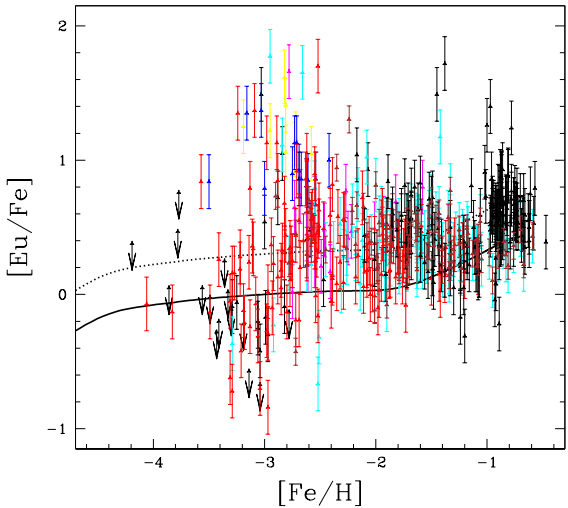,height=3in}
\end{center}
\caption{ {\bf Eu evolution with a time delay distribution.} Same as in Fig.~\ref{fig:EuFeSFR1} (SFR1 case) for a distribution of coalescence timescales corresponding to Fig.~\ref{fig:delay} (see text for more details). Solid and dotted curves correspond to two NSM fractions, of 0.002 and 0.004, respectively.
}
\label{fig:combinatoire}
\end{figure}

When fitting the observed early cosmic evolution of Eu in the NSM scenario, the coalescence timescale $\Delta t_\mathrm{NSM}$ is constrained to be short, typically in the range $50$--$200$~Myr depending on the SFR (see~\S~\ref{sec:standard}), as binary systems with longer timescales are not merging at large redshifts. As discussed in Sect.~\ref{sec:mergerdelay},  such systems with a relatively low value of $\Delta t_\mathrm{NSM}$ represent a significant fraction of all the systems, but not all of them. In Fig.~\ref{fig:combinatoire}, we show the evolution of Eu using a broader, and more realistic, distribution of coalescence timescales. Based on the distribution plotted in Fig.~\ref{fig:delay}, we assume 
13\% of the NSM in the $0$--$10$~Myr range ($\Delta t_\mathrm{NSM}=5$~Myr), 
10\% in the $10$--$100$~Myr range ($\Delta t_\mathrm{NSM}=30$~Myr), 
15\% in the $100$--$500$~Myr range ($\Delta t_\mathrm{NSM}=300$~Myr),
and
62\% with a coalescence timescale above 500~Myr ($\Delta t_\mathrm{NSM}=1$~Gyr). The calculation is carried out for the SFR1 mode and we present the result for both  $\alpha = 0.002$ (solid line) and $\alpha= 0.004$ (dotted line).  When compared to Fig.~\ref{fig:EuFeSFR1}, the early evolution is slightly modified due to the contribution of systems with even shorter coalescence timescales than in our reference case. As mergers are rare events with large Eu yields (compared to CCSN which are more frequent but produce less Eu), a large dispersion can be expected at low metallicity when only a small number of mergers affected the local environment.  This could explain the observed dispersion shown in  Fig.~\ref{fig:EuH} and discussed in  Sect.~\ref{sect_obs}.

\section{The predicted NSM rate }
\label{sec:mergerrates}

\begin{table*}
\caption{\label{tab:eventrate}
Predictions for the merger rate within the horizon of advanced Virgo/LIGO 
obtained with the constraint from the cosmic evolution of Eu using SFR1, SFR2, SFR3 models.
For each SFR, the lower and higher values of the predicted rate correspond to the higher and lower limits on the yield of Eu in mergers, i.e. $2 \times 10^{-4}$~M$_\odot$ and $7 \times 10^{-5}$~M$_\odot$, respectively. The range of independent predictions compiled by \citet{abadie10} is given on the first line for comparison.
Following \citet{abadie10}, we adopt 200 Mpc (resp. 420 Mpc) for the size of the horizon for NS-NS (resp. NS-BH) mergers.}

\begin{center}
\begin{tabular}{ccc}
& NS-NS merger rate ($\mathrm{yr^{-1}}$)
& NS-BH merger rate ($\mathrm{yr^{-1}}$) \\
\hline 
\citet{abadie10}
& $40$ ($0.4$--$400$)
& $10$ ($0.2$--$300$) \\
\hline
SFR1  & 2.4 -- 6.7 & 2.7 -- 7.7\\
 \hline
SFR2 & 2. -- 5.7&2.3 -- 6.87 \\
\hline
SFR3 &3.8 -- 10.9 &4.3 -- 12.4 \\
\end{tabular}
\end{center}
\end{table*}

\begin{table*}
\caption{\label{tab:kn}
Predictions for the kilonova rate for three different surveys: Palomar Transient Factory (PTF, limiting magnitude of $21$ and area of 
$2700\, \mathrm{deg^2}$, \citealt{law10}), Large Synoptic Survey Telescope (LSST, limiting magnitude  of $24.5$, area of $18000\, \mathrm{deg^2}$, \citealt{abell09}) and Euclid (limiting magnitude of $24.5$, area of $15000\, \mathrm{deg^2}$, \citealt{laureijs11}) for SFR1, SFR2, SFR3  models. 
For each SFR, we consider either the most pessimistic case ('low') or the most optimistic case ('high') corresponding to the highest and lowest possibles values of the yield of Eu, as in Table~\ref{tab:eventrate}. For each case, the range of predicted rates corresponds to the lowest and highest values of the peak luminosity of a kilonova (see text).
}

\begin{center}
\begin{tabular}{cccc}
& \multicolumn{3}{c}{Kilonova rate ($\mathrm{yr^{-1}}$)}\\
& PTF & LSST & Euclid\\
\hline 
SFR1 low &0.0018 --  0.034 &1.4 -- 22.9 & 1.2 --  19.1 \\
SFR1 high & 0.005 -- 0.096& 4.1 --  65.4& 3.3 --  54.5 \\
\hline
SFR2 low &0.0014 --  0.028 &1.2 -- 20.1 & 1.0 --  16.7 \\
SFR2 high & 0.004 -- 0.08& 3.5 --  57.4& 2.9 --  47.9 \\
\hline
SFR3 low & 0.003 -- 0.054 &2.3 -- 36.4 &  1.9 -- 30.3\\
SFR3 high & 0.008 --  0.16 &6.6 --  103.9 &  5.5 --  86.6\\
\end{tabular}
\end{center}

\end{table*}


\begin{figure}[htb!]
\begin{center}
\epsfig{file=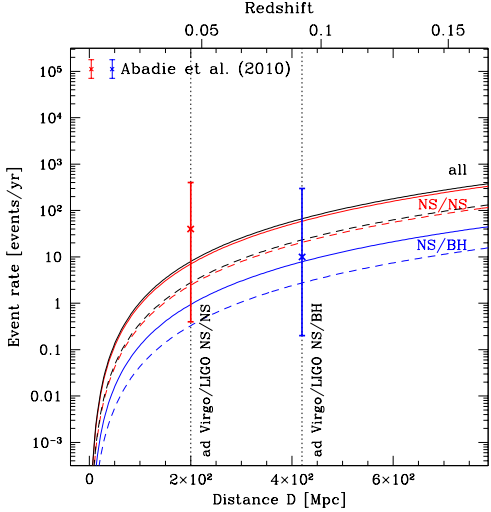, width=0.9\linewidth}
\end{center}
\caption{
The integrated merger rate (NS-NS, red line, NS-BH, blue line, and total, black line) as a function of the distance (lower axis) and the redshift (upper axis). Two vertical lines indicate the size of the horizon of the gravitational wave detectors advanced Virgo/LIGO. For comparison, the  central value and range of the predicted rates within this horizon compiled by \citet{abadie10} is also indicated for both types of mergers.
The calculation corresponds to the SFR1 mode with the standard prescription (solid line: Eu yield of
$7 \times 10^{-5}$~M$_\odot$ and binary fraction $\alpha = 0.002$) and with a modified prescription (dashed line: 
$2 \times 10^{-4}$~M$_\odot$ and binary fraction $\alpha = 0.0007$), both leading to exactly the same  cosmic evolution of Eu plotted in Fig.~\ref{fig:EuH}.
\label{fig:eventrate}
}
\end{figure}

\begin{figure}[htb!]
\begin{center}
\epsfig{file=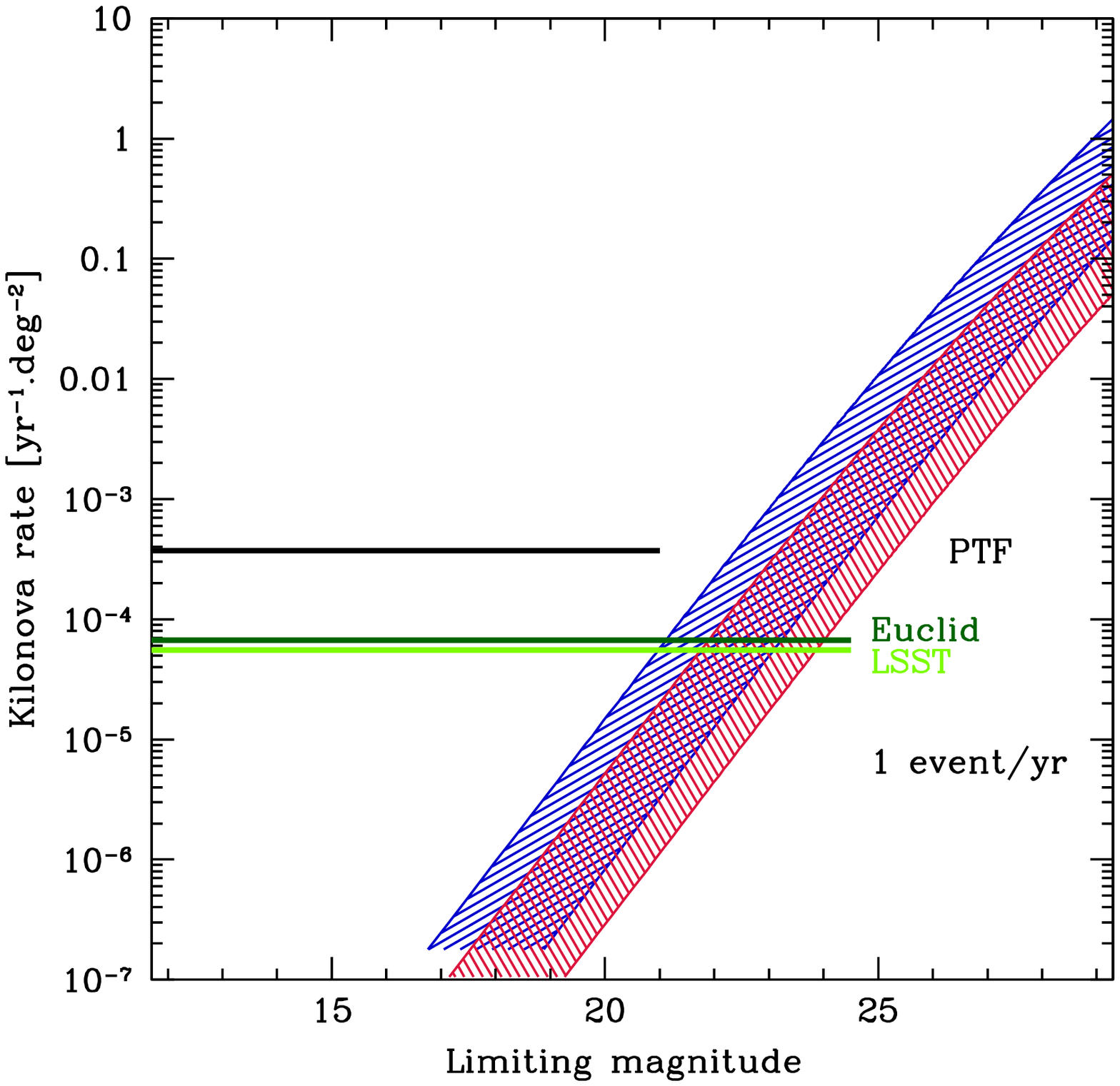, width=0.9\linewidth}
\end{center}
\caption{
The integrated kilonova rate per $\mathrm{deg^2}$ 
as a function of the limiting magnitude for the two same cases as in Fig.~\ref{fig:eventrate}, i.e. SFR1 with the standard prescription (blue) or the modified prescription (red) for the Eu yield. The width of the shaded area in each case takes into account the theoretical uncertainties on the peak luminosity of the kilonova (see Eq.~(\ref{eq:kn})).
For comparison, horizontal yellow and green lines indicate the limit of 1 event per year for three surveys (PTF, LSST and Euclid, see Sect.~\ref{sec:kn} and Table~\ref{tab:kn} for the limiting magnitude and area of each survey). 
\label{fig:kn}
}
\end{figure}

Once the fraction $\alpha$ and the coalescence timescale $\Delta t_\mathrm{NSM}$ have been adjusted to reproduce the cosmic evolution of Eu, as described in the previous section, the cosmic evolution of the merger rate is determined as all free parameters entering in Eq.~(\ref{eq:Rnsm}) are known. The result  
is plotted as a function of the redshift in Fig.~\ref{fig:explosions} for the SFR1 and SFR3 modes, together with the CCSN rate and the NS and BH birth rates.
The SFR evolution is shown for comparison.
We also plot the local merger rate taken from the compilation by
 \citet{abadie10}.

\subsection{Merger rate in the horizon of gravitational wave detectors}

On the basis of the cosmic evolution of the merger rate, it is possible to estimate the rate expected in the horizon of the gravitational wave detectors advanced Virgo and ad LIGO and to compare those rates with independent predictions, as compiled by \citet{abadie10}.
The result is plotted in Fig.~\ref{fig:eventrate} for the SFR1 scenario. 
When considering the production of Eu in the NSM scenario, both parameters, the Eu yield and the fraction of binary mergers $\alpha$, are degenerate when adjusted on the cosmic evolution of Eu.
This is not the case for the calculation of the event rates. Therefore, we
present in 
Fig.~\ref{fig:eventrate}
the results for two sets of parameters: 
either our standard Eu yield of  $7 \times 10^{-5}~M_\odot$ and a binary fraction $\alpha= 0.002$ (solid lines), 
or a larger Eu yield of $2 \times 10^{-4}~M_\odot$ (as found in NSM simulations, see Sect.~\ref{sect_euy}) and a lower binary fraction $\alpha= 0.0007$ (dashed lines), both leading to exactly the same cosmic evolution of Eu.
The predicted rates are given in Table~\ref{tab:eventrate} for the two extreme sets of parameters.

Interestingly, the integrated merger rate in the horizon of gravitational wave detectors obtained in the present study using our standard value for the Eu yield from NSM  ($7 \times 10^{-5}~M_\odot$)
is close to the central value of the predictions compiled by \citet{abadie10}. In contrast the upper value of the Eu yield tends to disfavor the upper half of the predicted range given by \citet{abadie10}, as clearly illustrated in Fig.~\ref{fig:eventrate}.

\subsection{Kilonova rate}
\label{sec:kn}

As discussed in the introduction, NS-NS/BH mergers can potentially have two main types of electromagnetic counterparts, short GRBs and kilonovae. Short GRBs are associated to the production of an ultra-relativistic outflow along the polar axis of the post-merger BH-torus system, with large uncertainties on the beaming angle of the jet. Therefore, the prediction of the short GRB rate from the merger rate requires a specific treatment  \citep[see e.g.][]{guetta06, wanderman15, Fryer12}, which we do not include in the present study. On the other hand, kilonovae are associated with a quasi-isotropic non-relativistic ejection that should be present in all systems. The luminosity of the kilonovae is powered by the radioactive decay of freshly synthesized elements and is consequently directly related to the nucleosynthesis of r-process nuclei. Note that NS/NS mergers are better sources of electromagnetic emission, as
the ejected mass is very low in a noticeable fraction of NS/BH mergers (see Fig. 7 in \cite{Belczynski08bis}). As NS/BH mergers are about one
order of magnitude less frequent than NS/NS mergers (see, Fig.~\ref{fig:eventrate}), this
does not affect our results.

Therefore, the detection of a kilonova would represent a major step to probe the contribution of NS-NS/BH mergers to the synthesis of heavy elements in the Universe \citep[see also][]{met15}. There is only one or two tentative detections to date. The first candidate is in association with the short GRB 130603B \citep{berger13,tanvir13}. However, 
this observation is not detailed enough (only a single photometric measurement) to prove firmly 
its kilonova nature. 
The second candidate is associated to GRB 060614, whose short or long nature is ambiguous, but is also based on a photometric near-infrared bump only \citep{yang15}.
More detections, with a photometric and spectroscopic follow-up, will be required to progress further. 
The peak of the kilonova is expected in the near-infrared range a few days after the merger (it may be bluer in the case of NS/BH mergers, \citealt{tanaka14}). The long-term evolution of the ejecta, which is decelerated by the ambient medium, also leads to radio emission over weeks. This signal is not considered here though it may also become detectable by future radio surveys \citep{nakar11,piran13}.

Our model allows to constrain the expected rate of kilonovae for a given large survey as a function of the limit magnitude in the optical/near-infrared range.
Following \citet{metzger10}, the peak luminosity scales as
\begin{equation}
L_\mathrm{KN,peak}\simeq 5\times 10^{40}\, \mathrm{erg/s}\, \left(\frac{f}{10^{-6}}\right)\left(\frac{v}{0.1c}\right)^{1/2}\left(\frac{M_\mathrm{ej}}{10^{-2}\, M_\odot}\right)^{1/2}\, ,
\label{eq:kn}
\end{equation}
where $v$ and $M_\mathrm{ej}$ are the velocity and mass of the ejecta respectively and $f$ is the heating efficiency \citep{arnett82,li98,metzger10}.
Compared to  \citet{metzger10}, the normalization in Eq.~\ref{eq:kn} has been corrected for 
a more realistic opacity of the r-process elements, which is expected to be $\sim 10^2$ times larger compared to the opacity of iron group elements considered in previous studies \citep{kas13} \citep[see also][]{kas14}.
Using the extreme values obtained in the systematic study of the dynamics of the ejecta carried out by \citet{baus13}, this leads to $L_\mathrm{KN,peak}\simeq \left(2.0-14\right)\times 10^{40}\, \mathrm{erg/s}$ (see also \citealt{grossman14}).

Assuming that all kilonovae have similar peak luminosities $L_\mathrm{KN,peak}$, we can deduce from the merger rate the number of kilonovae per year above a given flux. This leads to the results plotted in Fig.~\ref{fig:kn}, where the rate of kilonovae per year and per square degree is plotted as a function of the limiting magnitude of a given survey, in the SFR1 case with the two same sets of parameters (Eu yield, compact binary fraction $\alpha$) as in Fig.~\ref{fig:eventrate}. 
The width of the shaded area takes into account the range of  $L_\mathrm{KN,peak}$ discussed above.
Horizontal lines indicate the limit above which at least one event per year is expected for different surveys, taking into account their magnitude limit and their field of view. The results are listed in Table~\ref{tab:kn} for each SFR and for the two extreme values of the yield of Eu.
It appears that the current Palomar Transient Factory (PTF) survey (limiting magnitude of $21$ and area of $2700\, \mathrm{deg^2}$)
fails to detect kilonovae,
whereas the predictions are more optimistic for future surveys like the Large Synoptic Survey Telescope (LSST, limiting magnitude  of $24.5$, area of $18000\, \mathrm{deg^2}$) and mostly Euclid (limiting magnitude of $24.5$, area of view of $15000\,\mathrm{deg^2}$) with a well adapted spectral range, for which typically $\sim 2 - 100$ events per year can be expected.


\section{Conclusions}
\label{sect_conc}

We have computed the predicted evolution of the cosmic rates of CCSN and NSM for three possible history of the SFR density, which are representative of the current uncertainties at high redshift and reproduce important observational constraints such as the cosmic evolution of iron or the Thomson optical depth to CMB. We have deduced the corresponding expected evolution of Eu, a typical r-process elements, considering that the astrophysical sites for the r-process are either CCSN or NSM. In the latter case, the comparison to observations is used to constraint two free parameters, the binary fraction of compact objects and the typical coalescence timescale in NS binary systems.
 We obtain the following results:

\begin{itemize}

\item[(i)] The r-process evolution using the NSM scenario as the main astrophysical site is in good agreement with observations, assuming that the early evolution is dominated by mergers of binary systems with a coalescence timescale of the order of $\Delta t_\mathrm{NSM}\sim 100$~Myr. Such mergers represent a significant fraction of all mergers according to recent estimations obtained with detailed population synthesis codes;

\item[(ii)] In this merger scenario, the explosive events responsible for the production of r-process elements are more rare than in the supernova
scenario, but each event yields more Eu. Therefore, a larger scattering in the r-process abundance is expected in metal-poor stars enriched by mergers. We present calculations of the early evolution of Eu obtained when summing up contributions of mergers with a broad range of coalescence timescales according to predictions of population synthesis models and we find a reasonable  agreement with the level of scatter observed. At high metallicities, the cumulative effect leads to a low sensitivity with respect to the coalescence timescale;

\item[(iii)] The precise constraint on the dominant coalescence timescale relevant for the early evolution of Eu depends on the adopted SFR at high redshift, for which we have tested three scenarios representative of the current uncertainties, and on the iron yields of stars at low metallicity, for which we have compared two theoretical predictions made by either \citet{ww95} or \citet{Koba06}. The conclusion is that shorter coalescence timescales are favored by scenarios for which the production of iron at early times is more intense, i.e. for larger SFR at high redshift and larger iron yields coming from CCSN. In the standard conditions, we constrain the coalescence timescale $\Delta t_\mathrm{NSM}$ to be in the range of 50-200~Myr;

\item[(iv)] In the merger scenario which is favored by observations at low metallicity, two additional parameters are degenerate: the fraction $\alpha$ of NS in a binary system with another NS or a BH and the yield of Eu per merger. This second parameter is constrained to be in the range $7\times 10^{-5}$--$2\times 10^{-4}\, M_\odot$ per merger in the most recent nucleosynthesis calculations. Adopting these two values leads to an upper and a lower limit on $\alpha$ ($0.002-0.0007$) which constrain the merger rate as a function of the redshift in the Universe;

\item[(v)] We use the previous calculation to predict the expected rate of mergers within the horizon of gravitational wave detectors.
We find that the predicted rate as constrained by the cosmic evolution of Eu falls well within the range of predictions by other independent methods compiled by \citet{abadie10} but clearly favors the mid or lower values, and not the most optimistic predictions. We predict typically 2 to 10 NS-NS mergers per year and an equivalent number of NS-BH mergers per year;

\item[(vi)] Considering the most recent estimates of the opacity of r-process elements, we also predicted the rate of kilonovae in current and future optical/near-infrared large surveys. Our results are compatible with the lack of kilonova detections by PTF and are optimistic for the future LSST and Euclid projects, though with a large uncertainty: from 1 kilonova per year in the most pessimistic case to 100 events per year in the most optimistic case.

\end{itemize}

Our results can be improved in the future in several ways. Successful r-process in CCSN with well-defined Eu yields, including a possible dependence with metallicity, would be very useful to definitively check the consistency of this scenario with observations. A better determination of the SFR at large redshift and a convergence of theoretical predictions for iron yields in primordial CCSN would also strongly reduce the uncertainties on the parameters of the model in the merger scenario. In addition, more measurements of the abundance of r-process elements in stars at even lower metallicity ($\left[\mathrm{Fe}/\mathrm{H}\right]<-3.5$) would also allow to better constrain the model parameters. Finally, the predictions for the kilonova detection rate in future surveys could be improved by taking into account the correction for the spectral range of the instruments. Additional predictions for the rates of other transients associated with mergers, i.e. short gamma-ray bursts and radio flares, could also be deduced from the predicted merger rate to provide new diagnostics in the future.

\section*{Acknowledgements}

EV thanks warmly  Keith Olive, Jean-Pierre Lasota and Tsvi Piran
for constant and fruitful discussions.
The authors thank specifically
Gilles Esposito-Far\`ese for his help on the physics of mergers.
EV and FD acknowledge
 the French Program for High Energy astrophysics (PNHE) for financial support.
K.B. is sponsored by the NCN grant Sonata Bis2 ( DEC - 2012/07/ST9/01360) 
and by the polish Science Foundation "Master" subsidy.
This work is made in the ILP LABEX (under reference ANR-10-LABX-63) supported by French state funds managed by the ANR 
within the Investissements d'Avenir programme under reference ANR-11-IDEX-0004-02. 
SG is FRS-FNRS research associate.

\end{document}